\documentclass[prb,twocolumn]{revtex4-1}

\usepackage{amsfonts,amsmath,amsthm,amssymb}
\usepackage{mathrsfs}
\usepackage[mathscr]{eucal}
\usepackage{color}
\usepackage{graphicx}
\usepackage{adjustbox}
\usepackage{hyperref}
\definecolor{linkcolour}{rgb}{0,0.2,0.6}
\hypersetup{colorlinks,breaklinks,
            linkcolor=linkcolour,citecolor=linkcolour,
            filecolor=linkcolour, urlcolor=linkcolour}

\usepackage{bm}
\usepackage{cancel}
\usepackage{bbm}
\usepackage{placeins}
\usepackage{multirow}
\usepackage[table]{xcolor}

\usepackage[figuresright]{rotating}
\makeatletter
    \def\CT@@do@color{%
      \global\let\CT@do@color\relax
            \@tempdima\wd\z@
            \advance\@tempdima\@tempdimb
            \advance\@tempdima\@tempdimc
    \advance\@tempdimb\tabcolsep
    \advance\@tempdimc\tabcolsep
    \advance\@tempdima2\tabcolsep
            \kern-\@tempdimb
            \leaders\vrule
                    \hskip\@tempdima\@plus  1fill
            \kern-\@tempdimc
            \hskip-\wd\z@ \@plus -1fill }
\makeatother

\newcommand\ver{\mathcal{A}_\beta}
\newcommand\lr[1]{\left(#1\right)}
\newcommand\cint[1]{\frac{\de #1}{2\pi i}}
\newcommand\vd{{\vphantom{\dagger}}}

\newcommand\zed{\mathcal{Z}}
\newcommand\blam{{\bm{\lambda}}}
\newcommand\bmu{{\bm{\mu}}}
\newcommand\bgam{{\bm{\Gamma}}}
\newcommand\bsig{{\bm{\Sigma}}}
\newcommand\order[1]{\mathcal{O}\left(#1\right)}
\newcommand\de{\textrm{d}}
\newcommand\De{\textrm{D}}
\newcommand\eq[1]{\begin{equation}#1\end{equation}}
\newcommand\alg[1]{\begin{align}#1\end{align}}
\newcommand\tr[1]{\textrm{#1}}

\newcommand\ket[1]{\left|#1\right\rangle}

\newcommand\bra[1]{\left\langle#1\right|}
\newcommand\bbrakket[2]{\left\{#1\middle|#2\right\}}

\begin{document}

\title{On the Structure of Edge State Inner Products in the Fractional Quantum Hall Effect}
\author{R.~Fern}
\affiliation{Rudolf Peierls Centre for Theoretical Physics, Oxford University, 1 Keble Road, Oxford OX1 3NP, United Kingdom.}
\author{R.~Bondesan}
\affiliation{Rudolf Peierls Centre for Theoretical Physics, Oxford University, 1 Keble Road, Oxford OX1 3NP, United Kingdom.}
\author{S.~H.~Simon}
\affiliation{Rudolf Peierls Centre for Theoretical Physics, Oxford University, 1 Keble Road, Oxford OX1 3NP, United Kingdom.}

\begin{abstract}
We analyse the inner products of edge state wavefunctions in the fractional quantum Hall effect, specifically for the Laughlin and Moore-Read states.
We use an effective description for these inner products given by a large-$N$ expansion ansatz proposed in recent work by J.~Dubail, N.~Read and E.~Rezayi, PRB 86, 245310 (2012).
As noted by these authors, the terms in this ansatz can be constrained using symmetry, a procedure we perform to high orders.
We then check this conjecture by calculating the overlaps exactly for small system sizes and compare the numerics with our high-order expansion.
We find the effective description to be very accurate.
\end{abstract}

\pacs{?}

\maketitle

\section{Introduction}

The quantum Hall (qH) effect is one of the most influential and important phenomena in quantum condensed matter, having applications to quantum computing\cite{nayak2008non} and a deep underlying mathematical structure\cite{girvin1999quantum}.
On the analytical side, much of the progress has been made thanks to the close links between qH wavefunctions and conformal field theories (CFTs), which have been well documented and exploited in numerous works since the seminal 1991 paper by Moore and Read\cite{moore1991nonabelions, read1999beyond, li2008entanglement, dubail2012edge, hansson2017quantum}.
These trial wavefunctions can be expressed as chiral correlation functions of operators from particular CFTs and are excellent descriptions of exact qH states as found numerically\cite{laughlin1983anomalous,storni2010fractional,rezayi2009non}.

The edge states of qH wavefunctions are gapless excitations which are localised along the edge of the system\cite{wen1990chiral,chang2003chiral}.
If one considers the qH effect in a rotationally symmetric geometry (as we will in this work) within a radially increasing confining potential then the ground state is a circular droplet of uniform density and edge excitations can be thought of as chiral ripples around the circumference\cite{wen1992theory}.

However, calculating the overlaps of the resulting strongly correlated qH wavefunctions exactly is a very difficult problem, though one which is important for finding the energy and entanglement spectra of qH systems, as well as for computing observables.
Recent works have made progress by using the aforementioned description of wavefunctions in terms of CFT correlation functions\cite{dubail2012edge,wiegmann2005large}.
In this description trial wavefunctions are labelled by states from the CFT, which provides a far simpler space to work within.
We will closely follow the description in Ref.~\onlinecite{dubail2012edge} where the authors provide powerful constraints on the structure of the inner products of these trial states by appealing to CFT techniques as well as the underlying symmetries of qH states.

In Ref.~\onlinecite{dubail2012edge} the structure of these edge state inner products were used to analyse the particle and real-space entanglement spectra of qH states\cite{dubail2012real}.
However, the exact form for the inner product is a conjecture.
Therefore, even though this conjecture is extremely well-motivated, being based on an exact calculation for the trial wavefunctions of a $p_x+ip_y$ superconductor and having a strong physical basis, the authors do not provide a derivation of this result.
Specifically, the authors of Ref.~\onlinecite{dubail2012edge} use the ansatz that the inner product form is the exponential of a \textit{local} action, $S_N$, which is very difficult to prove in general, and we can only do so for the integer qH case, as demonstrated in a future publication\cite{usFUTUREinteger}.
Therefore it is important to further check the validity of this powerful result, and this constitutes the main outcomes of our work as we provide extensive numerical evidence in support of these claims.
We also use the constraints proposed in Ref.~\onlinecite{dubail2012edge} to calculate $S_N$ to very high orders.
In Ref.~\onlinecite{usFUTUREhamiltonian} we will build on these results to study effective Hamiltonians for the qH effect.

We begin in Sec.~\ref{Conformal Block Construction} with an overview of the construction of qH wavefunctions from CFT correlation functions.
We then discuss the general form for inner products in Sec.~\ref{Inner Products} and introduce the constraints provided by the symmetries of qH states in this new language.
This section also provides the high-order forms we calculate for $S_N$ in the Laughlin and Moore-Read cases.
Finally, we provide numerical evidence in support of the conjectures made by Ref.~\onlinecite{dubail2012edge} in Sec.~\ref{Numerical Evaluation}.

\section{Conformal Block Construction}
\label{Conformal Block Construction}

A nice review of the construction of qH states from conformal blocks is given in Ref.~\onlinecite{dubail2012edge} and we will cover those parts pertinent to our own discussion below.
Some rudimentary knowledge of CFT is assumed though excellent references for the subject include Refs.~\onlinecite{francesco2012conformal,ginsparg1988applied,ribault2014conformal}.

\subsection{General Construction}

The wavefunctions for quantum Hall states can be written as correlation functions of operators from a chiral conformal field theory which is generally made up of two \textit{sectors}, $\tr{CFT}_\tr{U(1)}\otimes\tr{CFT}_\chi$.
In this way an individual particle at position $z=x+iy$ is represented by an operator, $\ver(z)$, which can be decomposed into two parts,
	\eq{\ver(z) = :e^{i\sqrt\beta\varphi(z)}:\chi(z)}
where this first term is the vertex operator of a Bose field, $\varphi(z)$, and lives in the U(1) \textit{charge sector}, CFT$_\tr{U(1)}$ (the notation $:X:$ denotes normal ordering of $X$).
Note that this gives the operator a U(1) charge of $\sqrt\beta$.
The second term, $\chi(z)$, is a field from the \textit{statistics sector}, $\tr{CFT}_\chi$.

In this way trial wavefunctions at filling fraction $\nu=1/\beta$ are can be written as a correlation function
	\eq{\Psi^{(N,\beta)}_{\bra{v}}(\bm{z}) = \bra{v}c_\beta^{N\vd}\ver^\vd(z_1)\cdots\ver^\vd(z_N)\ket{0} \label{statedef}}
where $\bm{z}=\{z_1,\hdots,z_N\}$ are the positions of the $N$ particles making up the system, the factor $c_\beta^N$ is the background charge, which is required to give the correlator net U(1) charge of zero, and $\ket{0}$ is the vacuum state of the CFT with zero charge.
The out state, $\bra{v}$, is then used to form an individual edge excitation from the vacuum and is made up of the modes of the fields $\varphi(z)$ and $\chi(z)$.

Note that in this construction we have omitted the Gaussian factors usually attached to qH wavefunctions.
These will instead be included in the integration measure (see Eq.~\ref{integration measure}).
Furthermore, we will from now omit the $(N,\beta)$ label on the states, though this dependence remains implicit.

\subsection{The Laughlin State}

The simplest case is the Laughlin state, which has a trivial statistics sector ($\chi=\mathbbm{1}$) and so the CFT is simply that of the free boson\cite{francesco2012conformal}, $\tr{CFT}_{\tr{U(1)}}$.
The Bose field has a mode expansion of the form
	\eq{\varphi(z) = \varphi_0 - ia_0\ln(z) + i\sum_{n\neq0}\frac{a_n}{n}z^{-n} \label{bosonic mode expansion}}
where the individual modes satisfy
	\eq{[a_n,a_{-m}] = n\delta_{n,m}, \qquad [\varphi_0,a_0] = i}
and all other combinations commute.
We can phrase the background charge in terms of these modes as
	\eq{c_\beta^N = e^{-iN\sqrt\beta\varphi_0}}
and define the vacuum, $\ket{0}$, as the state which is annihilated by all modes $a_n$ for which $n\ge0$.

The Baker-Campbell-Hausdorff formula then gives
	\eq{\ver(z)\ver(w) = (z-w)^\beta:e^{i\sqrt\beta\lr{\varphi(z)+\varphi(w)}}:}
from which it is relatively straightforward to see that
	\eq{\Psi_{\bra{0}}(\bm{z}) = \prod_{i<j}(z_i-z_j)^\beta.}
This is the celebrated Laughlin wavefunction which describes a circular droplet of uniform quantum Hall fluid at filling fractions $\nu=1/\beta$ in a planar geometry.
In this geometry the power of an individual $z_i$ is the angular momentum of that $i^\tr{th}$ particle.

Edge modes are generated by applying the positive modes of the field to the out-state.
This adds angular momentum to the droplet and distorts the edge.
More specifically, if we define
	\eq{\bra{\blam} = \bra{0}\prod_{n\in\blam}a_n,\label{Laughlin_edgestates}}
where $\blam=\{\lambda_1,\lambda_2,\hdots\}$ is a semi-ordered set of positive integers ($\lambda_1\ge\lambda_2\ge\hdots>0$), then this produces the wavefunction
	\eq{\Psi_{\bra{\blam}}(\bm{z}) = P_\blam\prod_{i<j}(z_i-z_j)^\beta}
where these \textit{power sums}, $P_\blam$, have the form
	\eq{P_\blam = \prod_{n\in\blam}p_n,\qquad\qquad p_n=\sqrt\beta\sum_iz_i^n}
(NB: these $z_i$ are not normalised by factors of the droplet radius, $R$, as is sometimes the convention).
Recalling that the power on $z_i$ is the particle's angular momentum then the total angular momentum added, $\Delta L$, is equal to the degree of the polynomial, $|\blam|=\sum_i\lambda_i$.

\subsection{The Moore-Read State}

For the Moore-Read case, the statistics sector is the holomorphic sector of a free Majorana fermion CFT with field $\chi(z)=\psi(z)$.
This field has a mode expansion of the form
	\eq{\psi(z) = \sum_{n\in\mathbbm{Z}+\frac{1}{2}}\psi_nz^{-n-1/2}}
where the individual modes satisfy
	\eq{\{\psi_n,\psi_{-m}\} = \delta_{n,m}.}
The vacuum $\ket{0}$ is now also annihilated by the positive fermionic modes, $\psi_n$ for $n>0$ whilst the background charge remains unchanged.
Furthermore, given the parity symmetry of the free fermion theory, the correlator of an odd number of fermionic fields will always vanish, which leads to a slightly different picture when the particle number is odd.
We will initially focus on the case where $N$ is even.

Given that the OPE of the fermionic field is
	\eq{\psi(z)\psi(w) \sim \frac{1}{z-w},}
we pick up an extra factor of a \textit{Pfaffian} which was absent in the Laughlin case, giving a ground state of the form
	\eq{\Psi_{\bra{0}}(\bm{z}) = \tr{Pf}\lr{\frac{1}{z_i-z_j}}\prod_{i<j}(z_i-z_j)^\beta.}
where the Pfaffian, $\tr{Pf}(\cdots)$, is an antisymmetrised sum over all products of the fractions $\frac{1}{z_i-z_j}$,
	\eq{\tr{Pf}\lr{\frac{1}{z_i-z_j}} = \mathbbm{A}\lr{\frac{1}{z_1-z_2}
				\frac{1}{z_3-z_4}\cdots\frac{1}{z_{N-1}-z_N}}}
where $\mathbbm{A}$ denotes the operation of antisymmetrisation.
This is the Moore-Read (Pfaffian) wavefunction\cite{moore1991nonabelions} which it is believed is realised in graphene-based qH systems\cite{zibrov2016robust}.

As in the Laughlin case, edge excitations are generated by acting with the positive modes of our fields on the vacuum, adding angular momentum in the process.
The only change is that there are now two branches; the bosonic modes which are exactly as in the Laughlin case and an extra fermionic branch.
However, we must be careful about both parity symmetry and fermionic exclusion.
Therefore, denoting each out state as
	\eq{\bra{\blam\;;\;\bmu} = \bra{0}\prod_{n\in\blam}a_n\prod_{l\in\bmu}\psi_l
			\label{MR_edgestates}}
the set $\bmu=\{\mu_1,\mu_2,\hdots\}$ is an ordered set ($\mu_1>\mu_2>\hdots$) of positive half-integers ($\mu_i\in\mathbbm{Z}^{\ge0}+\frac{1}{2}$) with no repetitions ($\mu_i\neq\mu_j~~\forall i\neq j$) and an even number of elements.

The picture is subtly different when $N$ is odd.
In that case we must include an odd number of fermionic modes in our out state to ensure the total number of fermion fields within the correlator is even.
As such, the most compressed (lowest angular momentum) wavefunction in this case includes an extra, neutral fermion on the edge, and is defined by the out state $\bra{0}\psi_{1/2}$.
The structure of edge states is then similar to the even case except that $\bmu$ contains an odd number of elements.

\section{Inner Products}
\label{Inner Products}

In this section we will introduce the notation and general structure of inner products of edge state wavefunctions in the quantum Hall effect.
We will then review the conjecture and symmetry analysis of Ref.~\onlinecite{dubail2012edge} before using these symmetries to constrain the form of this inner product as much as possible.

\subsection{CFT Formulation}

We use curly Dirac notation for states in the physical Bargmann-Fock space of holomorphic polynomials to distinguish them from the states in the conformal field theory, i.e,
	\eq{\bbrakket{\bm{z}}{\Psi_{\bra{v}}} = \Psi_{\bra{v}}(\bm{z}).}
We are interested in finding the overlaps of such states, as defined by the following integration measure,
	\alg{\bbrakket{\Psi_{\bra{v}}}{\Psi_{\bra{w}}} & = \int_{\mathbbm{C}^N}
			\De\bm{z}\;\:\Psi_{\bra{w}}\Psi_{\bra{v}}^*, \\
		\De\bm{z} & = \frac{1}{\zed_{N,\beta}}\prod_{i=1}^N\lr{e^{-\frac{|z_i|^2}{2\ell_B^2}}\de^2z_i}, \label{integration measure}}
where $\ell_B$ is the magnetic length and $\zed_{N,\beta}$ is a normalisation constant defined such that
	\eq{\bbrakket{\Psi_{\bra{0}}}{\Psi_{\bra{0}}} = 1.}

Using our previous definitions of qH states in terms of conformal blocks we may now re-write this inner product in the Hilbert space of the CFT as
\begin{widetext}
	\eq{\bbrakket{\Psi_{\bra{v}}}{\Psi_{\bra{w}}} = \int\De\bm{z}\bra{w}c_\beta^N\ver(z_1)\cdots\ver(z_N)\ket{0}\bra{0}
				\ver^\dagger(\bar z_N)\cdots\ver^\dagger(\bar z_1)\lr{c_\beta^N}^\dagger\ket{v}.}
Therefore, we can define the inner product, $G_N$, within the space of the CFT as
	\eq{G_N = \int\De\bm{z}\;c_\beta^N\ver(z_1)\cdots\ver(z_N)\ket{0}\bra{0}
				\ver^\dagger(\bar z_N)\cdots\ver^\dagger(\bar z_1)\lr{c_\beta^N}^\dagger \label{GNdefinition}}
\end{widetext}
which produces a mapping from an inner product in the physical space into the CFT language of the form
	\eq{\bbrakket{\Psi_{\bra{v}}}{\Psi_{\bra{w}}} = \bra{w}G_N\ket{v}. \label{inner_equivalence}}
If we could evaluate Eq.~\ref{GNdefinition} we would then have the exact form for overlaps of qH edge states.
Unfortunately this proves to be an extremely difficult problem.

Instead we consider what form $G_N$ could have.
By definition, $G_N$ is a Hermitian and positive definite operator so we can write it in the form
	\eq{G_N = R^{2L_0}e^{-S_N} \label{metric_form}}
where $R$ is the radius of the qH droplet, $L_0$ is the zero mode of the Virasoro algebra (see below for its role in this problem) and $S_N=S_N^\dagger$.
This is the form proposed by Ref.~\onlinecite{dubail2012edge}, in which the authors argue that $S_N$ is a local action on the boundary of the droplet; it is a massive perturbation to the full CFT.

By itself, Eq.~\ref{metric_form} is just a definition of some operator $S_N$ onto which we shift our ignorance about the exact form of $G_N$.
However, the crucial simplification lies with the claim that $S_N$ is local.
This is a conjecture based on exact calculations for the trial states of $p_x+ip_y$ superconductors and on the assumption of charge screening within the bulk of the droplet and is one of the major constraints we imposed upon $S_N$ which allows one to make progress.
However, as it is not a rigorous statement it is important that we check its validity.
This, as well as computing its explicit form, is the main reason for our work with Sec.~\ref{Numerical Evaluation} providing numerical evidence in support of this ansatz.

\subsection{The Symmetries of $S_N$}

\subsubsection{Locality}

We begin with a short discussion on the conjecture of locality made by Ref.~\onlinecite{dubail2012edge}.
This claim is motivated partly by the fact that, for trial states of a $p_x+ip_y$ superconductor, the inner product operator equivalent to $G_N$ can be calculated exactly and it has the form claimed in Eq.~\ref{metric_form} for a local $S_N$.
Another motivation is provided by the \textit{generalised screening hypothesis}, the idea that all correlation functions within the bulk of the droplet should be short-range.

For the Laughlin state, this screening can be understood in terms of the plasma analogy, which maps the normalisation of the wavefunction, $\zed_{N,\beta}$, to a two-dimensional plasma in the presence of a background charge at finite temperature.
Numerically one finds that such a system is in a screening phase for $\beta<\beta_c$ where $\beta_c\simeq70$, meaning that correlation functions decay exponentially\cite{caillol1982monte,girvin1999quantum}.
A similar (though much more in-depth) analysis is also possible for the Moore-Read state with a similar conclusion\cite{bonderson2011plasma}.
As such, we expect that all the important physics in at least these two systems should be due to local interactions\cite{read2009non}.

This idea of screening suggests that the degrees of freedom in the system will not interact significantly with degrees of freedom at another location.
Given that the relevant degrees of freedom for the overlaps of edge states (which are localised on the boundary) will be located on the edge we surmise that the action will exist only the boundary.
Furthermore, it can include only local terms which do not generate interactions between well-separated regions along the edge.
As such, one may take an ansatz for $S_N$ which is local, of the form
	\eq{S_N = \sum_a\tilde s_a\int\de x\;\tilde\Phi_a(x) \label{action edge variable}}
where $x$ is the coordinate along the circumference of the droplet at radius $R$ (i.e, at $z=Re^{ix/R}$) and we sum over all possible \textit{local} operators, $\tilde\Phi_a$, which have a priori unknown coupling coefficients $\tilde s_a$.

It is important to understand which terms in this sum remain relevant as the system size increases.
The variation of each term becomes clear when we perform a change of variables from the circle, $x$, to the plane, $z$,
Under this substitution
	\eq{\tilde\Phi_a(x)=\lr{\frac{iz}{R}}^{d_a}\tilde\Phi_a(z) = \frac{1}{R^{d_a}}\Phi_a(z)}
where the second equality defines a new operator $\Phi_a(z) = (iz)^{d_a}\tilde\Phi_a(z)$.
Note that $d_a$ is the scaling dimension of the field $\Phi_a$ and recall that $R=\ell_B\sqrt{2\beta N}$ is the radius of the droplet where $\ell_B$ is the magnetic length.
This maps the action to
	\eq{S_N = \sum_a\frac{2\pi\tilde s_a}{R^{d_a-1}}\oint\cint{z}z^{-1}\Phi_a(z).}
Having re-phrased the action we note that the contour integral can be freely deformed and is a scale invariant quantity.
Based on the previous heuristics, we also expect the coefficients to vary as $\tilde s_a\sim\ell_B^{d_a-1}$ where $\ell_B$ is roughly the UV cut-off in the theory.
As such, we define dimensionless coefficients $s_a$ which we expect to be of order unity via $2\pi\tilde s_a = s_a(\ell_B\sqrt{2\beta})^{d_a-1}$ such that
	\eq{S_N = \sum_a\frac{s_a}{\sqrt{N}^{d_a-1}}\oint\cint{z}z^{-1}\Phi_a(z).}
Therefore, we see that the action is an expansion in $1/\sqrt{N}$ and so, for $N$ large enough, we can restrict our attention to only those operators where $d_a$ is small.

\subsubsection{Number Conservation}

A further observation is that the particle number in the Laughlin state is conserved\cite{dubail2012edge}.
This is a simple statement that the scalar product between any two states must be zero if the number of particles is not the same.
The operator which counts the particle number is $a_0/\sqrt\beta$ where $a_0$ is the zero mode of the Bose field and counts the total charge.
Therefore this constraint can be imposed by $[a_0,S_N]=0$.

As such, number conservation precludes any terms in $S_N$ which contain $\varphi_0$ and this imposes a strong constraint on the U(1) sector.
It prevents the inclusion of any vertex operators, $:e^{i\alpha\varphi(z)}:$, and requires any mention of the bosonic field to be as a derivative, i.e, the current $i\partial\varphi(z)$ or a descendant of it.
This does not imply any constraint on terms arising from the statistics sector.

\subsubsection{Rotational Invariance}

The penultimate constraint as a result of rotational invariance\cite{dubail2012edge}.
Quantum Hall states have well-defined angular momentum and the scalar product of any two excited states with different angular momentum must be zero.
The operator which measures angular momentum in the CFT language is $L_0$, the zero mode of the Virasoro algebra for the full CFT (i.e, $L_0=L_0^\tr{U(1)}+L_0^\chi$ contains a contribution from both the charge and statistics sectors).
Therefore rotational invariance forces us to conclude that $[L_0,S_N]=0$.

This constraint happens to be rather powerful.
So far the structure of $S_N$ is one of polynomials of the fields and their descendants multiplied by any a priori arbitrary function of $z$.
For example, in the Laughlin case all terms have the form
	\eq{\Phi_a(z) = f(z)\lr{i\partial^{\vphantom{2}}\varphi(z)}^{m_1}\lr{i\partial^2\varphi(z)}^{m_2}\cdots}
where the $m_j$ are integers which give the operator's scaling dimension as $d_a=\sum_jjm_j$.
This function $f(z)$ is then an a priori completely arbitrary function.
However, if we impose rotational invariance then this constrains it to be simply $f(z)=z^{d_a}$.
An analogous result exists for any other trial qH state constructed in the way we have discussed.

\subsubsection{Translational Invariance}

The final constraint is perhaps the most powerful as it allows many of the coefficients in the proposed expansion of $S_N$ to be fixed.
It is a consequence of the translational invariance of the Laughlin wavefunction and leads to the conclusion that the action must satisfy
	\eq{\left[e^{-S_N},a_{-1}\right] = \frac{1}{N\sqrt\beta}L_{-1}e^{-S_N} \label{transinv}}
where $L_{-1}$ is the generator of translations in our Virasoro algebra\cite{dubail2012edge}.
This allows us to fix the coefficients of all operators which have some non-trivial commutation relation with the $a_{-1}$ mode of the U(1) field.

It should be noted that in the notation of Ref.~\onlinecite{dubail2012edge} the relation equivalent to Eq.~\ref{transinv} is one in which $L_{-1}$ is replaced by $\tilde L_{-1}$, which is exactly the original Virasoro mode but given instead in terms of a shifted field, defined as
	\eq{\tilde\varphi(z) = \varphi(z) + iN\sqrt\beta \ln(z).}
This redefinition shifts the zero mode to $\tilde a_0=a_0-N\sqrt\beta$.
Such a shift is convenient as each state in their formulation includes the background charge $\bra{\tilde v}=\bra{v}c_\beta^N$.
Therefore, each state has a U(1) charge, $a_0\ket{v} = N\sqrt\beta\ket{v}$ which is cancelled by making this shift.

However, in our notation this shift is not necessary.
We include the background charge in the correlation function and so each of our edge states are automatically U(1) charge neutral.
However, the two notations are exactly equivalent, being related by a conjugation by $c_\beta^N$,
	\eq{c_\beta^N\tilde\varphi(z)c_\beta^{-N} = \varphi(z).}

To derive Eq.~\ref{transinv} we will analyse how translational symmetry can help us make exact statements about the inner products of edge states containing $p_1$ edge modes (where these modes correspond physically to a small shift of the whole droplet).
Consider the inner product,
	\eq{\bbrakket{\Psi_{\bra{v}a_1}}{\Psi_{\bra{w}}} = \bra{w}G_Na_{-1}\ket{v} = \int\De\bm{z}\bar p_1\bar\Psi_{\bra{v}}\Psi_{\bra{w}}. \label{a1byparts}}
Within this integral we can use the usual procedure for projection to the lowest Landau level\cite{girvin1984formalism} in which
	\eq{\bar z_i \to 2\ell_B^2\partial_i. \label{LLLProjection}}
This is as a result of the exponential factors hidden in $\De\bm{z}$.
We use these replace $\bar z_i$ with some derivative acting on the exponential and then integrate by parts, using that the wavefunction decays at infinity.
In this way $\bar p_1$ is replaced by
	\eq{\bar p_1 = \sqrt\beta\sum_{i=1}^N2\ell_B^2\partial_i}
which acts only on our holomorphic state, $\Psi_{\bra{w}}$.

We may then derive a Ward identity for translational invariance.
To do so we want to find some CFT operator $\mathcal{D}$ which reproduces the action of a derivative on the correlation function.
I.e, recalling that $\Psi_{\bra{w}}$ is a correlation function whose form is given by Eq.~\ref{statedef} we want to find some $\mathcal{D}$ satisfying
	\alg{\sum_i\partial_i\bra{w}c_\beta^N\ver(z_1)&\cdots\ver(z_N)\ket{0} = 
		\nonumber\\&\bra{w}\mathcal{D}c_\beta^N\ver(z_1)\cdots\ver(z_N)\ket{0}.}
In this way we map the translation operator into the CFT language.

In order to reverse-engineer the form of $\mathcal{D}$ we note that the -1$^\tr{st}$ mode of the Virasoro algebra, $L_{-1}$, is (almost) exactly the operator required, as it has the action
	\eq{[L_{-1},\mathcal{O}(z)] = \partial\mathcal{O}(z)}
on any operator $\mathcal{O}$ in the CFT.
As such, given that $L_{-1}\ket{0}=0$, the differentiation can be reproduced by
	\alg{\sum_i\partial_i\bra{w}c_\beta^N\ver(z_1)&\cdots\ver(z_N)\ket{0} = 
		\nonumber\\&\bra{w}c_\beta^NL_{-1}\ver(z_1)\cdots\ver(z_N)\ket{0}.}
We must then conjugate the Virasoro mode by the background charge to find $\mathcal{D}$.

To do so, we note that the Virasoro mode comes in two parts, $L_{-1}=L_{-1}^\tr{U(1)}+L_{-1}^{\vphantom{U(1)}\chi}$, where the latter, from the statistics sector, commutes with the background charge.
One then uses that the mode in the charge sector has the form,
	\eq{L_{-1}^\tr{U(1)} = \frac{1}{2}\oint\cint{z}:\lr{i\partial\varphi(z)}^2:.}
When this is commuted through the background charge it acquires an extra factor of $N\sqrt\beta a_{-1}$, leading to the conclusion that
	\eq{\sum_i\partial_i\Psi_{\bra{w}} = \Psi_{\bra{w}(N\sqrt\beta a_{-1} + L_{-1})}.}

Now that we have the action of $\bar p_1$ on our states we can reconsider Eq.~\ref{a1byparts} to see that
	\eq{\bbrakket{\Psi_{\bra{v}a_1}}{\Psi_{\bra{w}}} = 2\ell_B^2\sqrt\beta
			\bbrakket{\Psi_{\bra{v}}}{\Psi_{\bra{w}\lr{N\sqrt\beta a_{-1}+L_{-1}}}}}
As such we conclude that
	\alg{\bra{w}G_Na_{-1}\ket{v} & = 2\ell_B^2\sqrt\beta\bra{w}\left(N\sqrt\beta a_{-1} + L_{-1}\right)G_N\ket{v}}
which it is simple to show is equivalent to Eq.~\ref{transinv} upon the removal of the $R^{2L_0}$ factor in $G_N$.

\subsection{The Form of $S_N$}

\subsubsection{The Laughlin State}

We begin by summarising the consequences of the first three conditions for operators allowed in $S_N$ from the U(1) sector.
A complete basis which satisfies all these three conditions is given by
	\eq{T_\bgam = \oint\cint{z}z^{|\bgam|-1}:\prod_{n\in\bgam}i\partial^n\varphi(z):}
where $\bgam=\{\Gamma_1,\Gamma_2,\hdots\}$ is a semi-ordered set of positive integers, $\Gamma_1\ge\Gamma_2\ge\hdots>0$, of weight $|\bgam|=\sum\Gamma_i$.
For example,
	\eq{T_{11} = \oint\cint{z}z:\lr{i\partial\varphi(z)}^2: = 2L_0.}
The scaling dimension of each of these terms is $d_\bgam = |\bgam|$.

It should be noted that this basis is over-complete.
Certain partitions label what are effectively the same term given the freedom we have to integrate by parts within each $T_\bgam$.
For example, it is relatively simple to notice that $T_{21}=-T_{11}$.
We will omit such duplicates but this means that the coupling coefficient $s_\bgam$ will no longer appear only at order $\sqrt N^{|\bgam|-1}$, but also possibly at any order above this as well.
For example, if $T_{21}$ is simply $T_{11}$ then it is already clear that $T_{11}$ can appear at orders $\sqrt N^{-1}$ and $\sqrt N^{-2}$.

Therefore, we take the action to be some generic sum over this basis of terms,
	\eq{S_N = \sum_{\tr{unique } \bgam} \frac{s_\bgam}{\sqrt N^{|\bgam|-1}}T_\bgam}
where `unique $\bgam$' means that we omit any $\bgam$ which are duplicates via an integration by parts (for example, we include $T_{11}$ but not $T_{21}$).
Deciding what constitutes a unique $\bgam$ is discussed further in Appendix \ref{linear_ind_app}.
We then use the translational invariance constraint in Eq.~\ref{transinv} to constrain the vast majority of coefficients up to $6^\tr{th}$ order in $\sqrt N^{-1}$, with only two coefficients which cannot be determined by this method.
The final form is
	\alg{S_N = & -\frac{1}{6\sqrt\beta N}T_{111} + \frac{s_{22}}{N^{3/2}}T_{22} + \frac{1}{24\beta N^2}T_{1111} \nonumber\\
			& - \left(\frac{s_{22}}{2\sqrt\beta N^{5/2}}-\frac{1}{144\beta\sqrt\beta N^3}\right)(3T_{221}-2T_{111})
					\nonumber\\
			& + \frac{s_{33}}{N^{5/2}}T_{33} - \frac{1}{60\beta\sqrt\beta N^3}T_{11111}
					 + \order{N^{-7/2}} \label{action}}
where $s_{22}$ and $s_{33}$ are undetermined coupling constants which we expect to be of order unity, but with corrections of order $\sqrt N^{-1}$, i.e,
	\eq{\frac{s_{22}^{\vphantom{(}}}{N^{3/2}} = \frac{s_{22}^{(3)}}{N^{3/2}} + \frac{s_{22}^{(4)}}{N^2} + \hdots \label{s22 expansion}}
and similar for $s_{33}$.
We will now sketch a derivation of the form of Eq.~\ref{action}, carrying out the argument explicitly to 3rd order.

We begin by expanding $S_N$ in powers of $\sqrt N^{-1}$,
	\eq{S_N = S^{(0)} + \frac{S^{(1)}}{\sqrt N} + \frac{S^{(2)}}{N} + \cdots.}
To fix $S^{(0)}$ we simply note that the our ground state must be normalised.
This is the condition that
	\eq{\bra{0}R^{2L_0}e^{-S_N}\ket{0} = 1,}
and so $S^{(0)}\ket{0}=0$.
Given that the only term with a scaling dimension of zero is $T_1=a_0$ and $a_0$ will always evaluate to 0 (recall that we work with states $\ket{v}$ whose charge is zero) we can discard it and simply set $S^{(0)}=0$.

We can also quickly convince ourselves that $S^{(1)}=0$.
The only terms allowed at this order are $T_1=a_0=0$ and $T_{11}$, the latter of which has a non-trivial commutation with $a_{-1}$.
However, translational invariance, Eq.~\ref{transinv}, requires that the commutator of $S^{(1)}$ with $a_{-1}$ be zero, and so the coupling coefficient $s_{11}$ must be zero.

The first non-trivial term arises at order $N^{-1}$.
At this order our translational invariance condition reads
	\eq{[S^{(2)},a_{-1}] = -\frac{1}{\sqrt\beta}L_{-1}.}
By noting that
	\eq{[i\partial\varphi(z),a_{-1}] = z^{-2} \label{simple commutator}}
we see straight away that the term $T_{111}$ has commutation relation
	\eq{[T_{111},a_{-1}] = 3\oint\frac{\de z}{2\pi i}:\left[i\partial\varphi(z)\right]^2:\;=6L_{-1}.}
Therefore, we may conclude that $s_{111}=-\frac{1}{6\sqrt\beta}$ at this order whilst all other coefficients vanish.

At order $N^{-3/2}$ our translational invariance condition tells us that
	\eq{[S^{(3)},a_{-1}] = 0.}
The smallest set of non-vanishing linearly independent operators we are allowed at this order are $T_{11}, T_{111}, T_{1111}$ and $T_{22}$.
The first three have a non-trivial commutation relation with $a_{-1}$ so their coupling coefficients at this order must vanish.
$T_{22}$ however, commutes with $a_{-1}$ and so we can say nothing about its coupling coefficient.
As such,
	\eq{S_N = -\frac{1}{6\sqrt\beta N}T_{111} + \frac{s_{22}}{N^{3/2}}T_{22} + \order{N^{-2}}.}

Beyond these orders the picture becomes more involved but can still be approached in a manner similar to that presented here.
The only extra concern to consider is that $S_N$ is hermitian and is therefore made up only of hermitian operators.
As it transpires, some of the terms, $T_\bgam$, are neither hermitian nor can be made hermitian in combination with other $T_\bgam$.
The first casualty of this condition is $T_{222}$.

\subsubsection{The Moore-Read State}

In the Moore-Read case our (over-complete) basis of terms are labelled by two sets of integers,
	\eq{T_{\bgam,\bsig} = \oint\cint{z}z^{d_{\bgam,\bsig}-1}:\prod_{n\in\bgam}i\partial^n\varphi(z)\prod_{m\in\bsig}\partial^m\psi(z):}
where $\bgam$ is once again a semi-ordered set of positive integers but $\bsig$ is an ordered set of positive integers and zero, $0\le\Sigma_1<\Sigma_2<\hdots$ (i.e, no two $\Sigma_i$ are the same), with an even number of elements (due to parity symmetry).
We must also define an ordering for the product over $\bsig$ as $\psi(z)$ anti-commutes with itself.
We take
	\eq{\prod_{m\in\bsig}\partial^m\psi := \partial^{\Sigma_1}\psi\;\partial^{\Sigma_2}\psi\;\hdots.}
The scaling dimension of these terms is then $d_{\bgam,\bsig} = |\bgam|+|\bsig|+\ell(\bsig)/2$, where $\ell(\bsig)$ is the number of elements in $\bsig$.

As in the Laughlin case, we generate an action which is a sum over all the uniquely defined labels, $\bgam$ and $\bsig$ and constrain whatever terms we can by translational symmetry.
We find that the action has the form
	\alg{S_N =\; & \frac{s_{\emptyset,01}}{N^{1/2}}T_{\emptyset,01} - \frac{1}{6\sqrt\beta N}T_{111,\emptyset} \nonumber\\
		& + \frac{\lr{1-s_{\emptyset,01}/N^{1/2}+s_{\emptyset,01}^2/3N}}{2\sqrt\beta N}T_{1,01} 
				\nonumber\\
		& + \frac{s_{22,\emptyset}}{N^{3/2}}T_{22,\emptyset} + \frac{s_{\emptyset,12}}{N^{3/2}}T_{\emptyset,12} \nonumber\\
		& + \frac{1}{24\beta N^2}T_{1111,\emptyset} - \frac{1}{4\beta N^2}T_{11,01} \nonumber \\
		& + \frac{s_{3,01}}{N^2}\lr{T_{3,01}+3T_{2,01}} + \order{N^{-{5/2}}}
		\label{actionMR}}
where $s_{\emptyset,01}, s_{22,\emptyset}, s_{\emptyset,12}$ and $s_{3,01}$ are all undetermined constants assumed to be of order unity but with potential corrections of order $\sqrt N^{-1}$ etc.
Once again, we sketch a short proof of the lower orders of this result.

As was the case for the Laughlin state, we expand $S_N$ in powers of $\sqrt N^{-1}$ and apply the translational invariance constraint, Eq.~\ref{transinv}, order by order.
We first note that $S^{(0)}=0$ for exactly the reasons presented in the Laughlin calculation.
We then note that $S^{(1)}$ can be made up of only two non-trivial terms, namely $T_{11,\emptyset}$ and $T_{\emptyset,01}$. 
As we saw in the Laughlin case, the non-vanishing commutation relation which $T_{11,\emptyset}$ satisfies with $a_{-1}$ disqualifies it but the fermionic term, $T_{\emptyset,01}$ commutes with $a_{-1}$, and therefore might appear with any coefficient.
Therefore, we surmise that
	\eq{S^{(1)} = s_{\emptyset,01}T_{\emptyset,01}}
where $s_{\emptyset,01}$ is an unknown.
Therefore, this first term is, up to an overall factor, simply the stress-energy for the statistics sector,
	\eq{S^{(1)} = -2s_{\emptyset,01}\oint\cint{z}zT^\psi(z)}
where $T^\psi(z)$ is the holomorphic component of the stress-energy tensor.

At the next order our constraint is of the form
	\eq{\left[-S^{(2)} + \frac{\lr{S^{(1)}}^2}{2},a_{-1}\right] = \frac{1}{\sqrt\beta}L_{-1}.}
We have just seen that $S^{(1)}$ commutes with $a_{-1}$ so the contribution involving this vanishes whilst $S^{(2)}$ must produce $L_{-1}$.
The terms allowed by this order are those we saw at order $\sqrt N^{-1}$ and also
	\alg{T_{111,\emptyset} & = \oint\cint{z}z^2:\lr{i\partial\varphi(z)}^3:, \\
		 T_{1,01} & = \oint\cint{z} z^2:i\partial\varphi(z)\psi(z)\partial\psi(z):.}
Therefore, given that the generator of translations in the Virasoro algebra of the full CFT has the form
	\eq{L_{-1} = \frac{1}{2}\oint\cint{z}:\lr{\lr{i\partial\varphi(z)}^2 - \psi(z)\partial\psi(z)}:,}
and recalling the identity in Eq.~\ref{simple commutator}, it is straightforward to see that
	\eq{S^{(2)} = -\frac{1}{6\sqrt\beta}T_{111,\emptyset} + \frac{1}{2\sqrt\beta}T_{1,01}.}

Beyond this point the calculation once again progresses in a very similar manner with operators which commute with our $a_{-1}$ mode being assigned a priori unknown coefficients and the remaining terms being fixed by translational invariance.
We must also pay attention to hermiticity which first excludes $T_{2,01}$ appearing by itself as we might expect it to in $S^{(3)}$.

\section{Numerical Analysis}
\label{Numerical Evaluation}

The major aim of this paper is to test that the conjectured form for inner products proposed in Ref.~\onlinecite{dubail2012edge} agrees with the true overlaps of quantum Hall edge states.
We calculate these overlaps exactly by generating the edge states in a single-particle basis of monomials, making use of the description of these states in terms of Jack polynomials as pioneered in Refs.~\onlinecite{bernevig2008generalized,bernevig2008model,bernevig2008properties}.
We must then perform a basis transformation on these Jack polynomial states to produce the edge states generated by the CFT modes, a procedure which is well understood for the Laughlin case\cite{baratta2011jack} but must be considered on a case-by-case basis for Moore-Read states.

Unfortunately, this method limits us to quite modest values of $N$ as the Hilbert space dimension quickly becomes too large to store individual states.
For the Laughlin case this limits us to $N=12$ whilst for the Moore-Read case the size of the Hilbert space in this monomial basis is more limited and only grows too large for our methods above $N=18$.
However, given that the effects we are trying to observe are as small as $1/N^3$ it is crucial that our method is exact as the errors on a similar implementation with, for example, Monte Carlo would be unlikely to converge without enormous computer time.
Furthermore, the fact that $N$ is small makes the corrections we are looking for more clearly visible and makes the subsequent excellent agreement all the more impressive.

\subsection{The Laughlin State}

\subsubsection{Fitting the coefficients of $S_N$}

In order to evaluate the accuracy of the ansatz, Eq.~\ref{metric_form}, we must first find suitable fits for the coefficients we were unable to constrain by symmetry arguments alone.
To do this we minimise the Frobenius norm of the matrix corresponding to deviations between the exact data and this ansatz, $G_N(s_{22},s_{33})=R^{2L_0}e^{-S_N(s_{22},s_{33})}$ where $S_N$ is defined in Eq.~\ref{action}.
So, given a matrix of overlaps calculated exactly, $(O_N)_{i,j} = \bbrakket{\Psi_{\bra{i}}}{\Psi_{\bra{j}}}$, we set the values of the coefficients $s_{22}$ and $s_{33}$ such that the ``error'',
	\eq{e_N = \left\lVert G_N(s_{22},s_{33})-O_N\right\rVert_F,}
is minimised, where $\left\lVert X\right\rVert_F=\tr{Tr}\lr{XX^\dagger}$ denotes the Frobenius norm of the matrix $X$.

We expect this description to be most accurate for lower angular momentum states, as we discuss in the following subsection.
Therefore we restrict the basis of states $\ket{i}$ in which we perform this minimisation procedure to be only those in the $\Delta L=4$ subspace.
Furthermore, we work with states $\ket{i}$ which are normalised.
Given that the individual normalisations of the states (Eq.~\ref{Laughlin_edgestates}) vary quite significantly, this normalisation ensures that diagonal matrix elements of $O_N$ and $G_N$ are each close to unity.
This is important as the Frobenius norm only cares about the magnitude of the deviations, and so this normalisation ensures that each matrix element is weighted roughly equally during the minimisation procedure.

Recall that we truncate $G_N$ to be a sixth order expansion with errors of order $N^{-7/2}$.
Therefore, we should expect that $e_N$ is of the same order and this introduces some uncertainty in the values we fit for $s_{22}$ and $s_{33}$.
For example, because we a priori expect $s_{22}$ to appear as a third order contribution, any fit using our sixth order expansion of $G_N$ will have some error of order $N^{-2}$ as compared with an infinite-order expansion of $G_N$.
Similarly, our fit to $s_{33}$ will suffer errors of order $N^{-1}$.

We could, of course, reduce this error by expanding $G_N$ to higher orders, a procedure which is algorithmically simple and so could be delegated to a computer.
However, we have chosen to truncate at sixth order because the seventh order contribution includes three new coefficients which cannot be constrained by translational symmetry.
As such, if one continues to expand $G_N$ to higher and higher orders one is forced to accept more and more fit coefficients.
Whilst this would no doubt increase the accuracy of $G_N$, it somewhat diminishes the utility and results in over-fitting the data.

\begin{figure*}[h!]
	\centering
	\includegraphics[scale=0.43]{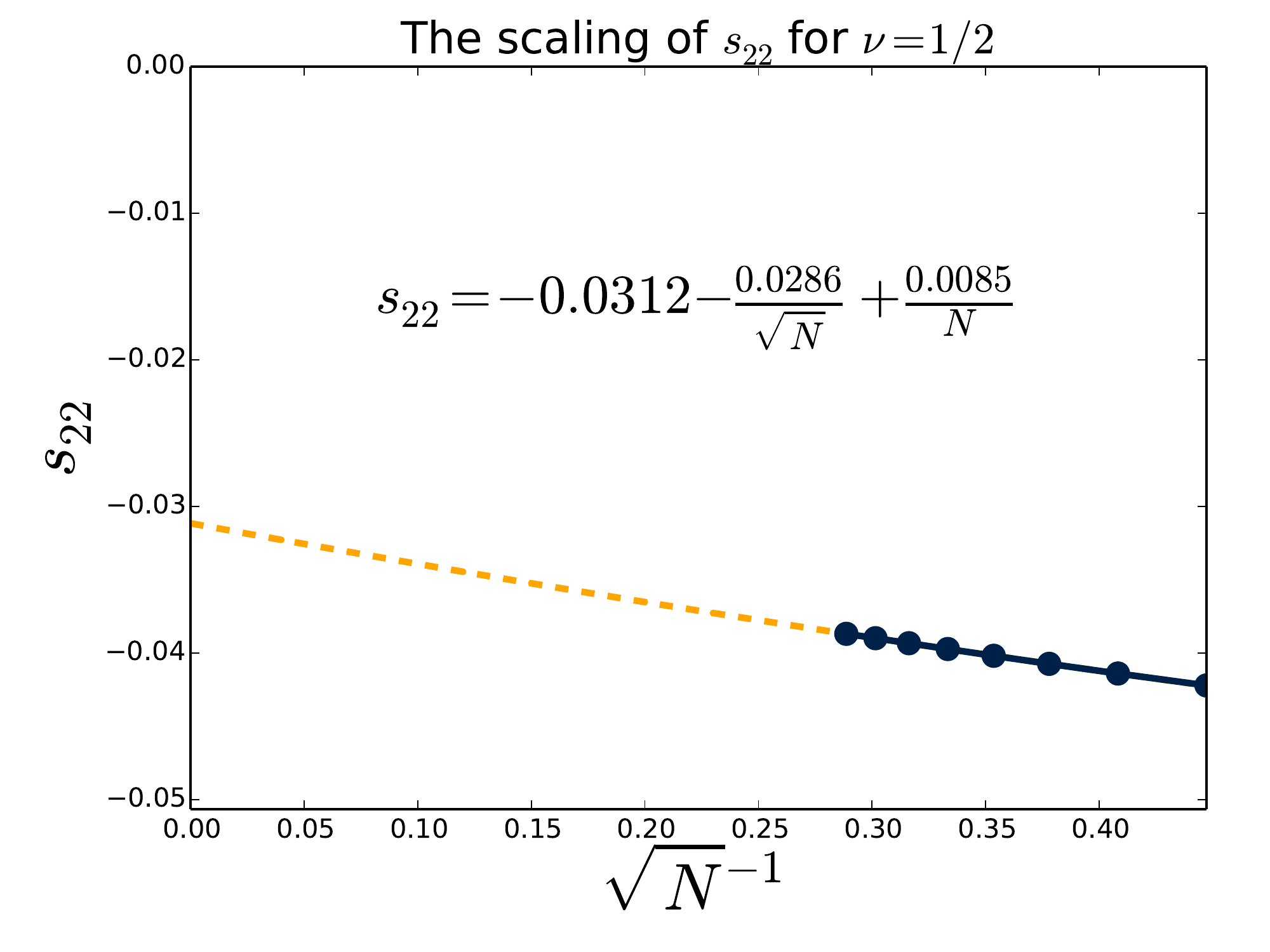}
	\includegraphics[scale=0.43]{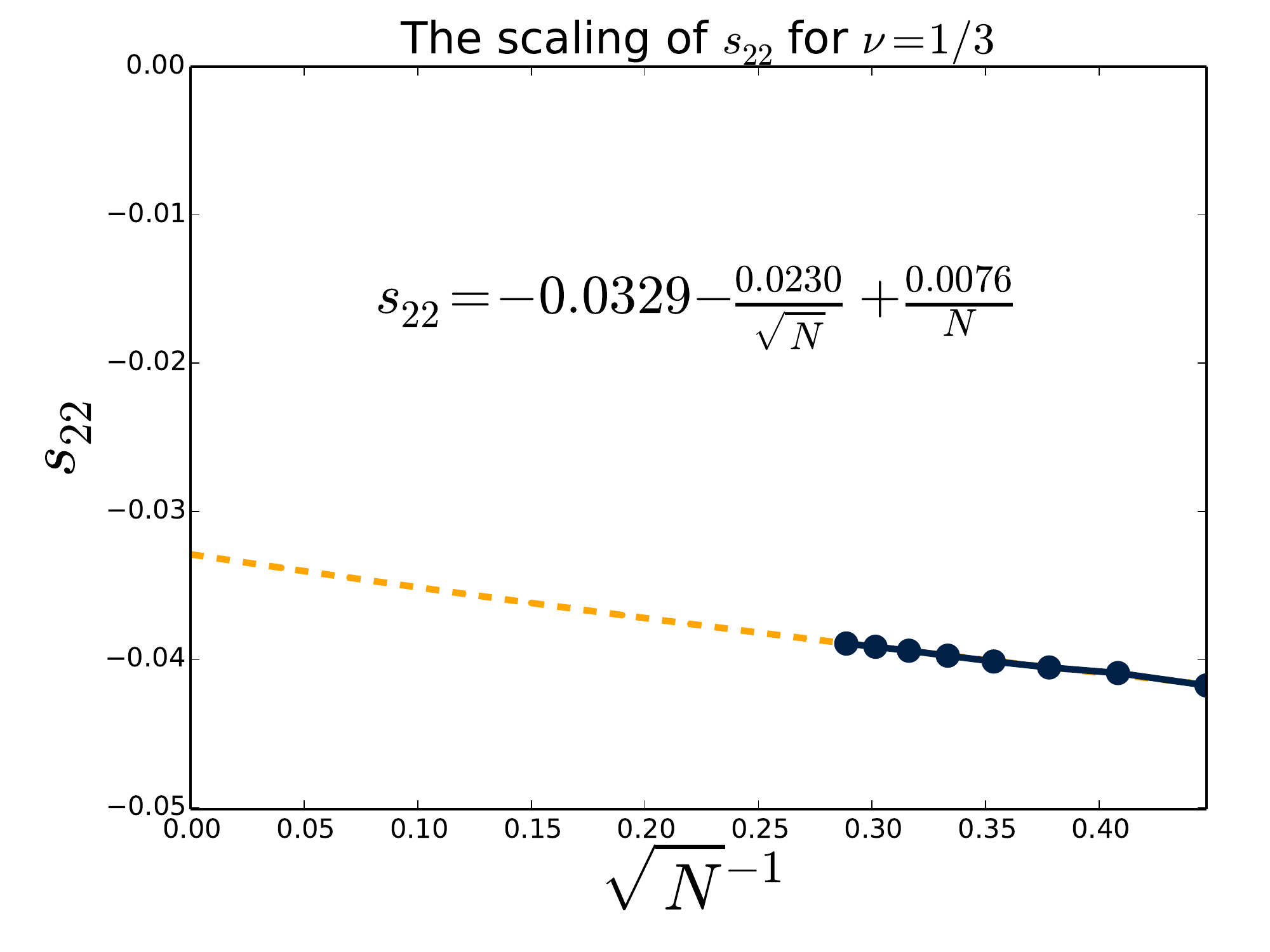}
	\includegraphics[scale=0.43]{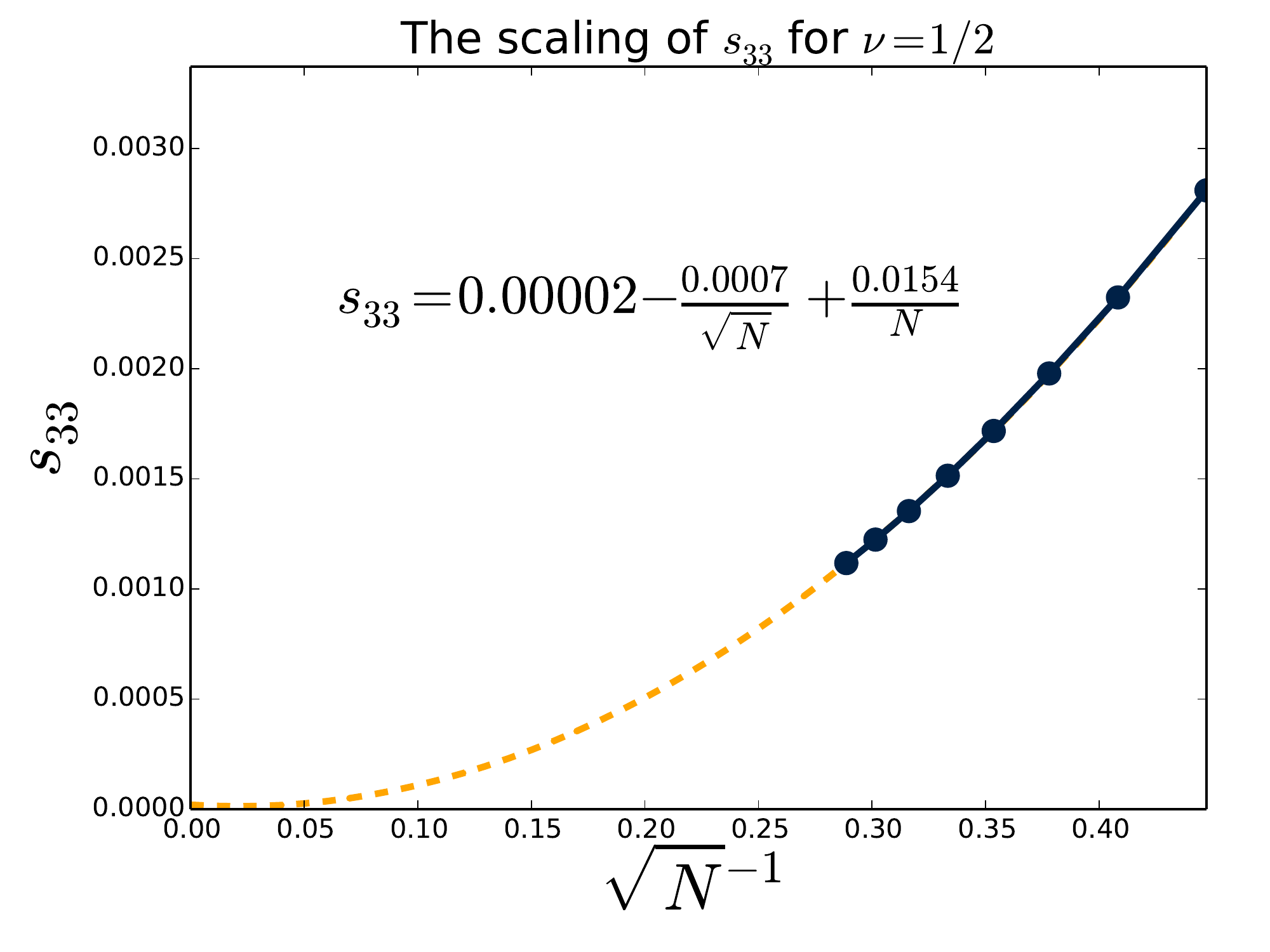}
	\includegraphics[scale=0.43]{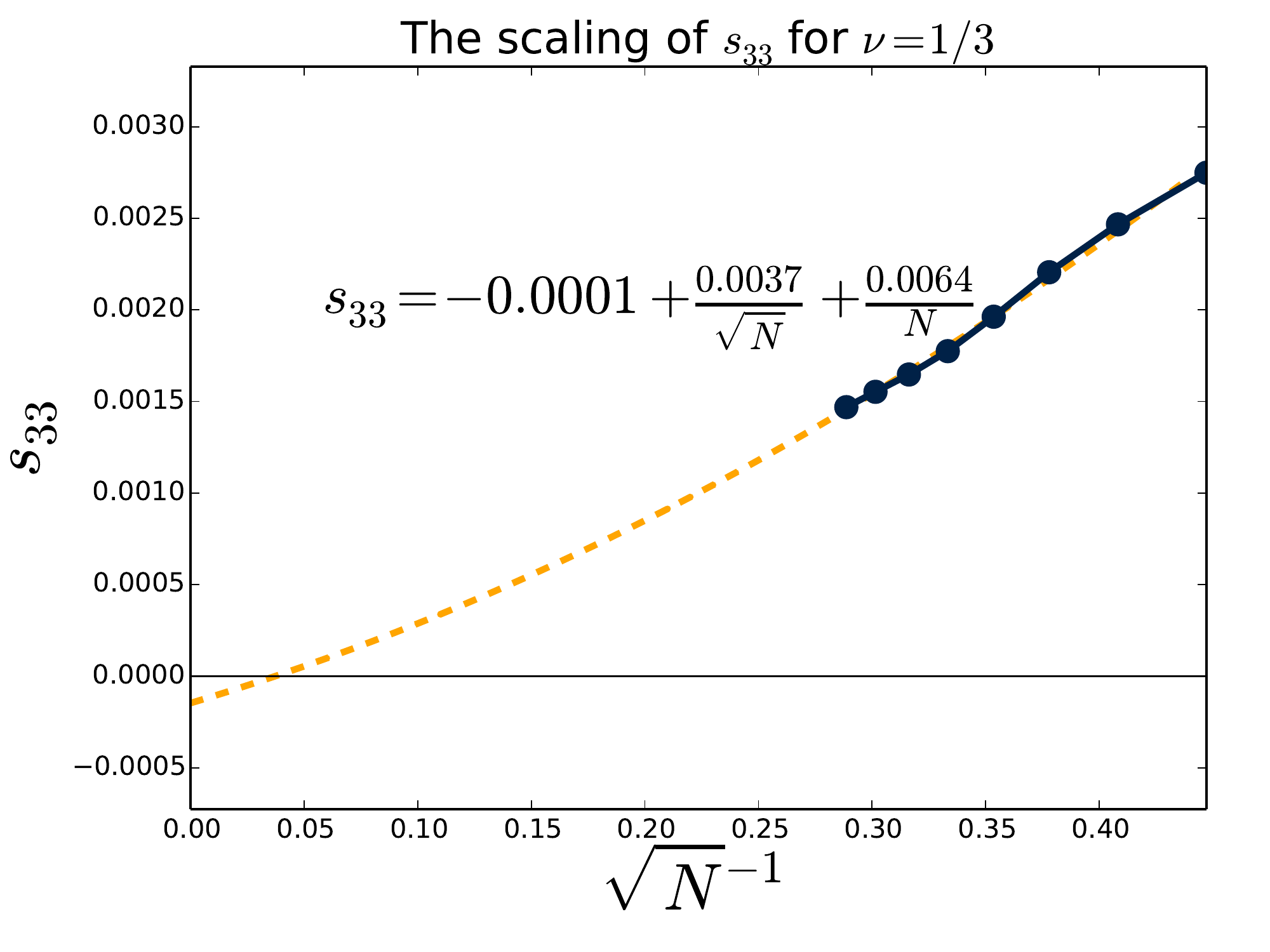}
	\caption{We calculate the values for the coupling coefficients $s_{22}$ and $s_{33}$ which best describe the data for the Laughlin state at filling $\nu=1/2$ and $\nu=1/3$ as shown by the dark blue points in the figures.
		We then perform a weighted least-squares fit of the variation of these coefficients with $\sqrt N^{-1}$ and find the forms given in each plot, as shown by the orange dashed curve.
		Although the points are calculated exactly, the weighting of each point assumes that the errors due to truncation on $s_{22}$ are of order $N^{-2}$ and of order $N^{-1}$ for $s_{33}$.
		Specifically the fit is obtained by minimising the sum $s = \sum_i\lr{y_i-f(x_i)}^2/e_i^2$ where $y_i$ are our data points, $f(x_i)$ our fit function with variable $x_i=\sqrt{N}^{-1}$ and $e_i$ this assumed ``error''.}
	\label{Laughlin fits}
\end{figure*}

Therefore, we simply fit the values of $s_{22}$ and $s_{33}$ such that $G_N$ as expanded to sixth order provides the optimal description of the data, $O_N$ (a selection of this data for the largest case of $N=12$ is provided in tables \ref{5_2 overlaps} and \ref{5_3 overlaps}).
We plot the results of these fits in Fig.~\ref{Laughlin fits} for filling $\nu=1/2$ and $\nu=1/3$.
Firstly we note that the scaling of $s_{22}$ appears to be as expected, tending towards some constant value in both cases.
This is strong evidence in support of the scaling arguments from Ref.~\onlinecite{dubail2012edge}.
The scaling of $s_{33}$ is less clear and suggests that the term $T_{33}$, to which this coefficient is associated, may appear at order $N^{-3}$ or higher, instead of the $N^{-5/2}$ we expect.
Nevertheless, this does not contradict the scaling arguments as we expect each coefficient to include such sub-leading corrections.
The only surprise is that the leading term may vanish.
As such, this is still supporting evidence that this scaling analysis is valid.

\begin{table*}[t!]
	\centering
	\begin{tabular}{| c || c | c | c | c | c | c | c |}
\hline $\nu = 1/2$ & $\Psi_{\bra{1,1,1,1,1}}$ & $\Psi_{\bra{2,1,1,1}}$ & $\Psi_{\bra{2,2,1}}$ & $\Psi_{\bra{3,1,1}}$ & $\Psi_{\bra{3,2}}$ & $\Psi_{\bra{4,1}}$ & $\Psi_{\bra{5}}$ \\ \hline\hline\multirow{2}{*}{$\Psi_{\bra{1,1,1,1,1}}$} & 120.0000\cellcolor{orange!20} & 7.0711 & 0.4167 & 0.4167 & 0.0246 & 0.0246 & 0.0014 \\ & \textbf{120.0000}\cellcolor{orange!20} & \textbf{7.0711} & \textbf{0.4167} & \textbf{0.4167} & \textbf{0.0246} & \textbf{0.0246} & \textbf{0.0000}\\ \hline\multirow{2}{*}{$\Psi_{\bra{2,1,1,1}}$} & 7.0711 & 12.2628\cellcolor{orange!20} & 1.4206 & 2.1187 & 0.1660 & 0.2482 & 0.0243 \\ & \textbf{7.0711} & \textbf{12.2626}\cellcolor{orange!20} & \textbf{1.4205} & \textbf{2.1185} & \textbf{0.1667} & \textbf{0.2500} & \textbf{0.0246}\\ \hline\multirow{2}{*}{$\Psi_{\bra{2,2,1}}$} & 0.4167 & 1.4206 & 8.0167\cellcolor{orange!20} & 0.2482 & 1.4073 & 0.9355 & 0.2699 \\ & \textbf{0.4167} & \textbf{1.4205} & \textbf{8.0177}\cellcolor{orange!20} & \textbf{0.2500} & \textbf{1.4072} & \textbf{0.9352} & \textbf{0.2778}\\ \hline\multirow{2}{*}{$\Psi_{\bra{3,1,1}}$} & 0.4167 & 2.1187 & 0.2482 & 6.0461\cellcolor{orange!20} & 0.3635 & 1.3812 & 0.2013 \\ & \textbf{0.4167} & \textbf{2.1185} & \textbf{0.2500} & \textbf{6.0472}\cellcolor{orange!20} & \textbf{0.3635} & \textbf{1.3805} & \textbf{0.2083}\\ \hline\multirow{2}{*}{$\Psi_{\bra{3,2}}$} & 0.0246 & 0.1660 & 1.4073 & 0.3635 & 5.9861\cellcolor{orange!20} & 0.2425 & 1.6569 \\ & \textbf{0.0246} & \textbf{0.1667} & \textbf{1.4072} & \textbf{0.3635} & \textbf{5.9940}\cellcolor{orange!20} & \textbf{0.2500} & \textbf{1.6528}\\ \hline\multirow{2}{*}{$\Psi_{\bra{4,1}}$} & 0.0246 & 0.2482 & 0.9355 & 1.3812 & 0.2425 & 3.8653\cellcolor{orange!20} & 1.0914 \\ & \textbf{0.0246} & \textbf{0.2500} & \textbf{0.9352} & \textbf{1.3805} & \textbf{0.2500} & \textbf{3.8709}\cellcolor{orange!20} & \textbf{1.0878}\\ \hline\multirow{2}{*}{$\Psi_{\bra{5}}$} & 0.0014 & 0.0243 & 0.2699 & 0.2013 & 1.6569 & 1.0914 & 4.3703\cellcolor{orange!20} \\ & \textbf{0.0000} & \textbf{0.0246} & \textbf{0.2778} & \textbf{0.2083} & \textbf{1.6528} & \textbf{1.0878} & \textbf{4.3855}\cellcolor{orange!20}\\ \hline
	\end{tabular}
	\caption{We show a table calculating the overlaps of Laughlin-type edge states numerically and with our expression $G_N$ at filling $\nu=1/2$ and system size $N=12$.
		The entries in the row $i$ and column $j$ correspond to $\bbrakket{\Psi_{\bra{i}}}{\Psi_{\bra{j}}}/R^{2\Delta L}$ where the upper entry is the value we find by taking exact overlaps numerically and the lower entry (in bold) is the value we find when we use $G_N$.}
	\label{5_2 overlaps}
\end{table*}

\begin{table*}[t!]
	\centering
	\begin{tabular}{| c || c | c | c | c | c | c | c |}
\hline $\nu = 1/3$ & $\Psi_{\bra{1,1,1,1,1}}$ & $\Psi_{\bra{2,1,1,1}}$ & $\Psi_{\bra{2,2,1}}$ & $\Psi_{\bra{3,1,1}}$ & $\Psi_{\bra{3,2}}$ & $\Psi_{\bra{4,1}}$ & $\Psi_{\bra{5}}$ \\ \hline\hline\multirow{2}{*}{$\Psi_{\bra{1,1,1,1,1}}$} & 120.0000\cellcolor{orange!20} & 5.7735 & 0.2778 & 0.2778 & 0.0134 & 0.0134 & 0.0006 \\ & \textbf{120.0000}\cellcolor{orange!20} & \textbf{5.7735} & \textbf{0.2778} & \textbf{0.2778} & \textbf{0.0134} & \textbf{0.0134} & \textbf{0.0000}\\ \hline\multirow{2}{*}{$\Psi_{\bra{2,1,1,1}}$} & 5.7735 & 12.1301\cellcolor{orange!20} & 1.1539 & 1.7241 & 0.1104 & 0.1653 & 0.0132 \\ & \textbf{5.7735} & \textbf{12.1299}\cellcolor{orange!20} & \textbf{1.1538} & \textbf{1.7240} & \textbf{0.1111} & \textbf{0.1667} & \textbf{0.0134}\\ \hline\multirow{2}{*}{$\Psi_{\bra{2,2,1}}$} & 0.2778 & 1.1539 & 7.9512\cellcolor{orange!20} & 0.1653 & 1.1423 & 0.7563 & 0.1794 \\ & \textbf{0.2778} & \textbf{1.1538} & \textbf{7.9520}\cellcolor{orange!20} & \textbf{0.1667} & \textbf{1.1421} & \textbf{0.7562} & \textbf{0.1852}\\ \hline\multirow{2}{*}{$\Psi_{\bra{3,1,1}}$} & 0.2778 & 1.7241 & 0.1653 & 5.9398\cellcolor{orange!20} & 0.2897 & 1.1193 & 0.1337 \\ & \textbf{0.2778} & \textbf{1.7240} & \textbf{0.1667} & \textbf{5.9407}\cellcolor{orange!20} & \textbf{0.2897} & \textbf{1.1189} & \textbf{0.1389}\\ \hline\multirow{2}{*}{$\Psi_{\bra{3,2}}$} & 0.0134 & 0.1104 & 1.1423 & 0.2897 & 5.8779\cellcolor{orange!20} & 0.1611 & 1.3393 \\ & \textbf{0.0134} & \textbf{0.1111} & \textbf{1.1421} & \textbf{0.2897} & \textbf{5.8836}\cellcolor{orange!20} & \textbf{0.1667} & \textbf{1.3371}\\ \hline\multirow{2}{*}{$\Psi_{\bra{4,1}}$} & 0.0134 & 0.1653 & 0.7563 & 1.1193 & 0.1611 & 3.7585\cellcolor{orange!20} & 0.8784 \\ & \textbf{0.0134} & \textbf{0.1667} & \textbf{0.7562} & \textbf{1.1189} & \textbf{0.1667} & \textbf{3.7629}\cellcolor{orange!20} & \textbf{0.8765}\\ \hline\multirow{2}{*}{$\Psi_{\bra{5}}$} & 0.0006 & 0.0132 & 0.1794 & 0.1337 & 1.3393 & 0.8784 & 4.2072\cellcolor{orange!20} \\ & \textbf{0.0000} & \textbf{0.0134} & \textbf{0.1852} & \textbf{0.1389} & \textbf{1.3371} & \textbf{0.8765} & \textbf{4.2189}\cellcolor{orange!20}\\ \hline
	\end{tabular}
	\caption{As in table \ref{5_2 overlaps} we show a comparison of exact calculations of Laughlin state overlaps found numerically (upper entry) with our approximation using $G_N$ (lower entry in bold) for $N=12$.
		This data is at filling $\nu=1/3$.}
	\label{5_3 overlaps}
\end{table*}

However, it is also interesting to note the strong similarity between the data for $\nu=1/2$ and $\nu=1/3$.
Firstly, the large-$N$ value of the coefficient, $s_{22}^{(3)}$, appears to be the same, or very similar, in both cases.
The sub-leading correction, $s_{22}^{(4)}$ also contains a remarkable coincidence.
For the integer qH effect at $\nu=1$ it is straightforward to show\cite{usFUTUREinteger} that this sub-leading contribution is $s_{22}^{(4)} = -1/24$ and so it is interesting to note that the values for these fractional cases are extremely close to $-1/24\sqrt\beta$ (where $\beta=1$ for the integer case).
In the $\nu=1/2$ case we find $s_{22}^{(4)}=-0.0286$ as compared with $-1/24\sqrt\beta=-0.0295$ and for the case at $\nu=1/3$ we have $s_{22}^{(4)}=-0.0230$ as compared with $-0.0241$.

For filling fractions $\nu=1/m$ where $m>3$ the structure is less clear.
Firstly, these states live in a larger single-particle Hilbert space as they have a larger total angular momentum, which makes storing the state difficult.
Secondly, we find that the convergence, even at comparable system sizes, is much worse, in much the same way that the data for $\nu=1/3$ is slightly more sporadic than that at $\nu=1/2$.
Nevertheless, we can perform the same analysis for $\nu=1/4$ and $\nu=1/5$ up to system sizes of $N=10$.
Whilst the resulting data is far noisier than that presented in Fig.~\ref{Laughlin fits}, a simple linear fit is potentially consistent with the cases above.
We find that $s_{22}^{(3)}=-0.030$ and $-0.027$ in the $\nu=1/4$ and $\nu=1/5$ states respectively, similar to the values we fit in Fig.~\ref{Laughlin fits}.
Furthermore, we find that the slopes are $s_{22}^{(4)}=-0.0206$ and $-0.0190$ as compared to our guess, $-1/24\sqrt\beta$, which evaluates to approximately $-0.0208$ and $0.0186$ respectively.

These coincidences, as well as the suggestion that the leading contributions to $s_{33}$ vanish in both cases, seem to suggest deeper structure in this form for the inner product which it might be possible to access analytically.
For example, it may be possible to derive further identities like Eq.~\ref{transinv} which encode other, non-obvious symmetries of the problem.
One might also be able to use the large-$N$ expansions of generating functions for the 2D Dyson gas presented in Ref.~\onlinecite{wiegmann2005large}.
Furthermore, it is conceivable that the Matrix Product State description of these states, as pioneered in Refs.~\onlinecite{zaletel2012exact, estienne2013matrix}, might also  be used to make exact statements about some or all of these coefficients.

\subsubsection{Accuracy of the $S_N$ expansion}

Most important is that the form of $G_N$ accurately describes the overlaps of qH edge states.
Therefore, we input the fits for $s_{22}$ and $s_{33}$ shown in Fig.~\ref{Laughlin fits} into the inner product operator, $G_N(s_{22},s_{33})$ and calculate the form of inner products of edge states, once again for filling fractions $\nu=1/2$ and $\nu=1/3$.
Some results for 12-particle systems are presented in Tables \ref{5_2 overlaps} and \ref{5_3 overlaps} where we show the overlaps matrices in the $\Delta L=5$ sub-space (where recall $\Delta L$ is the amount of angular momentum added by the edge excitation).
The inner product as calculated by $G_N$ agrees extremely well with the exact data for every overlap.

However, the data also shows that the agreement is worse for inner products involving large-$n$ modes, $a_n$ with one of the worst agreements being the calculation of the normalisation of the $\Psi_{\bra{0}a_5}$ state.
That the agreements become worse for larger $n$ modes is not unexpected.
Consider, for example, the operator $T_{mm}$ which we expect to appear at order $N^{-(2m-1)/2}$.
The expectation of this operator in the state $a_{-n}\ket{0}$ for large $n$ varies as
	\eq{\bra{0}a_n\lr{T_{mm}}a_{-n}\ket{0} \sim n^{2m-1}.}
As such, if such an operator were to appear in the action, $S_N$, then it's leading contribution to the normalisation of the state $\Psi_{\bra{0}a_n}$ would vary as $\sim s_{mm}(n/\sqrt N)^{2m-1}$ where $s_{mm}$ would be its associated coupling constant (which we would be unable to fix using translational symmetry).
Therefore, even though we have neglected all $T_{mm}$ for $m>3$ on the grounds that they are much more irrelevant in $1/\sqrt N$ than the terms we have kept, this still poses problems for calculating matrix elements of states including modes $a_n$ where $n$ is comparable to $\sqrt N$.

Finally, one may wonder to what extent we can reject the inclusion of non-local terms in the action.
As such, in Appendix \ref{non local} we consider allowing $S_N$ to also include the well known \textit{Benjamin-Ono} term\cite{abanov2005quantum,wiegmann2012nonlinear}, which we expect to be the least irrelevant non-local contribution.
While this term is expected to appear at order $N^{-1}$, a scaling analysis shows that its effect is vanishingly small and falls off as $\sqrt N^{-b}$ where $b$ is larger than the order of terms we neglect in the fit.
This suggests that its presence is very unlikely.

\subsection{The Moore-Read State}

\subsubsection{Fitting the coefficients of $S_N$}

\begin{figure*}[t!]
	\centering
	\includegraphics[scale=0.43]{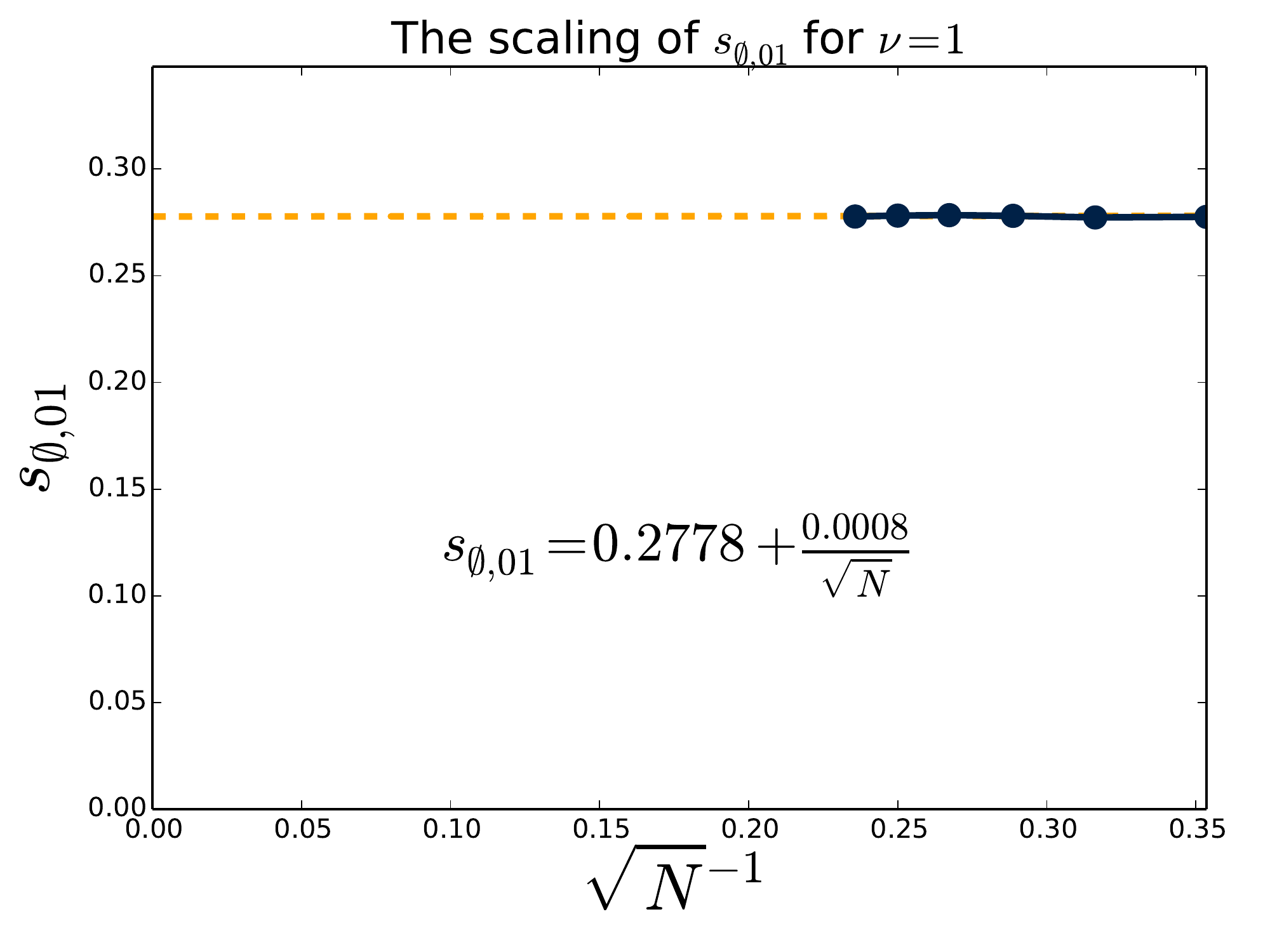}
	\includegraphics[scale=0.43]{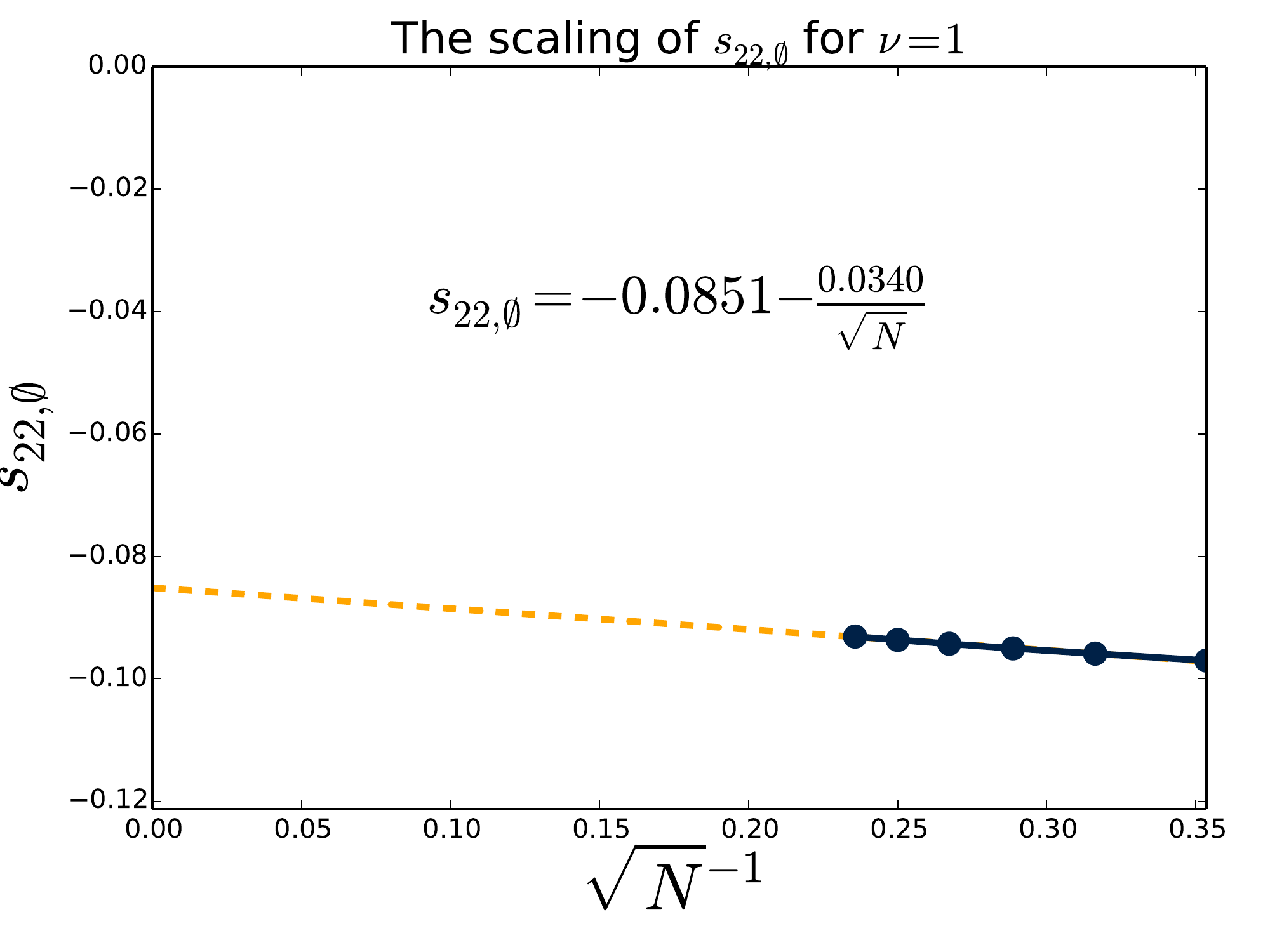}
	\includegraphics[scale=0.43]{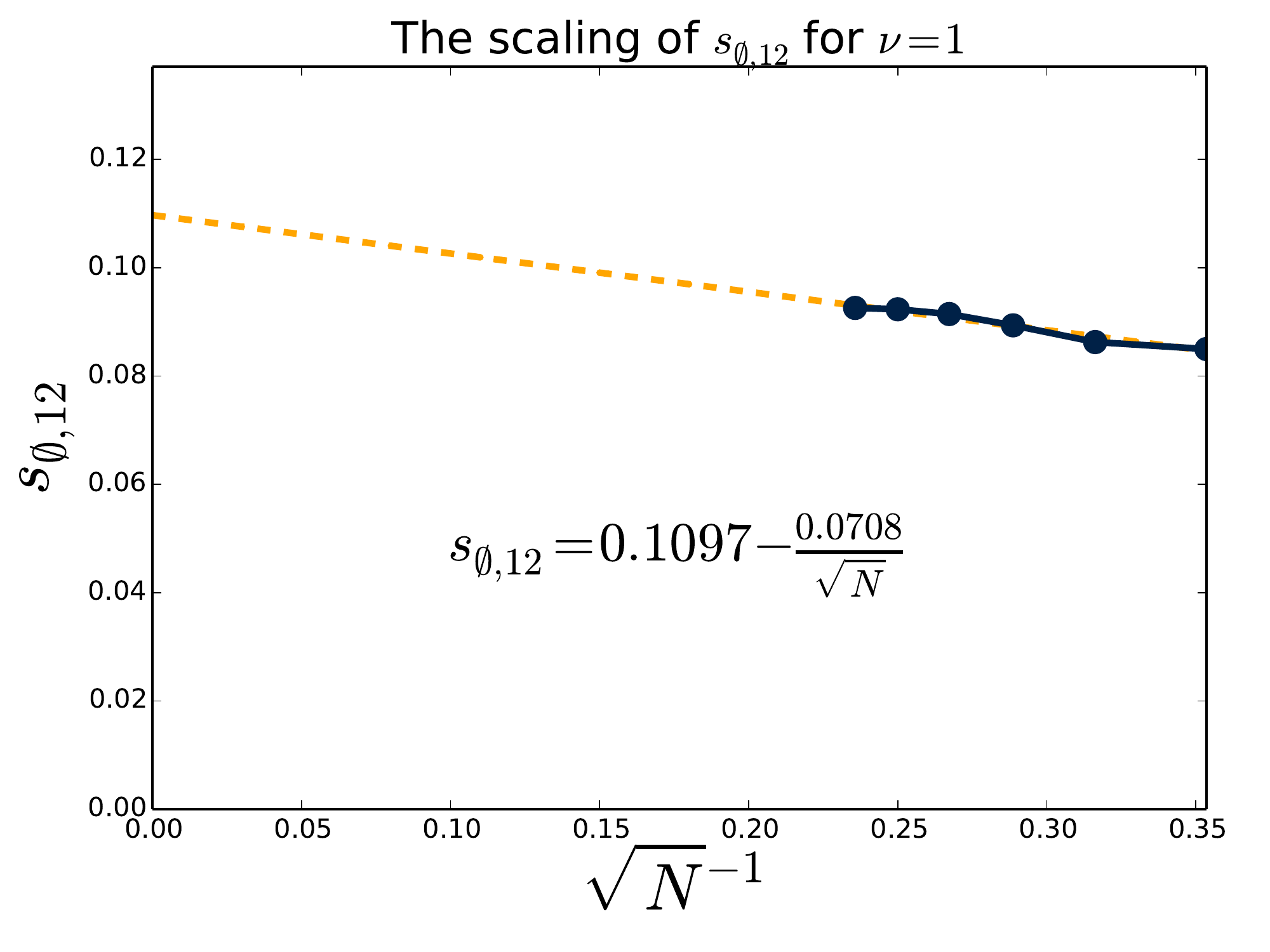}
	\includegraphics[scale=0.43]{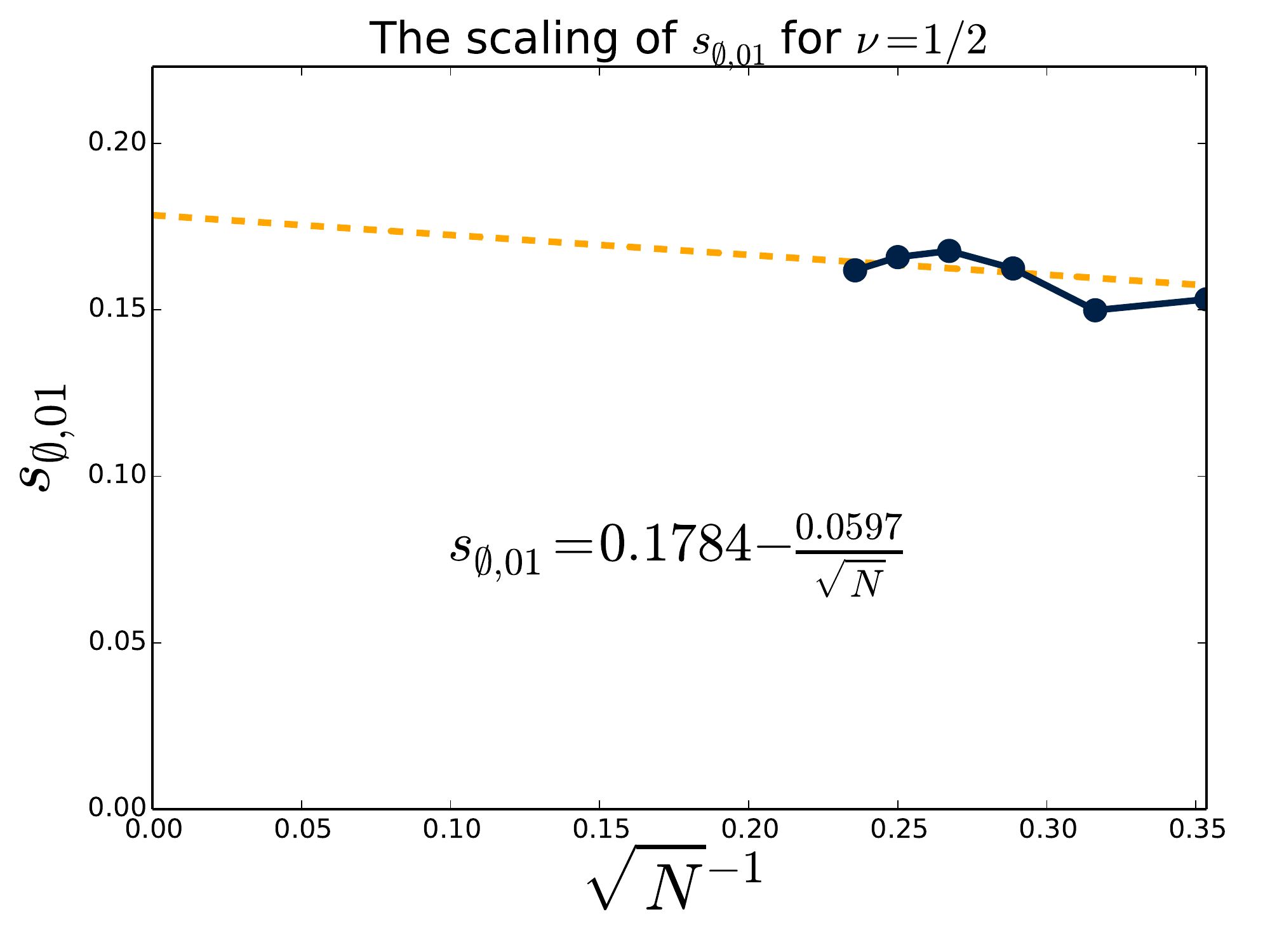}
	\caption{We calculate the values of coupling coefficients $s_{\emptyset,01}, s_{22,\emptyset}$ and $s_{\emptyset,12}$ which best describe the data for the Moore-Read edge states at filling $\nu=1$ and also for $s_{\emptyset,01}$ at filling $\nu=1/2$.
		These are shown by the dark blue points in the figures.
		We then perform a weighted least-squares fit of the variation of these coefficients with $\sqrt N^{-1}$ and find the forms given in each plot, as shown by the orange dashed curve.
		Each point is weighted according to the amount of error expected due to truncation.
		For example, we expect to calculate $s_{\emptyset,01}$ up to corrections of order $N^{-2}$ and so we set this as the error on each point and then minimise the sum of squared residuals weighted by this error as for the Laughlin case detailed in Fig.~\ref{Laughlin fits}.}
	\label{MR fits}
\end{figure*}

We would also like to analyse these claims for the Moore-Read state.
We proceed with fitting the coefficients of $S_N$ in exactly the same manner as for the Laughlin case, by minimising the Frobenius norm of deviations between the exact data and the results given by $G_N=R^{2L_0}e^{-S_N}$ where $S_N$ is defined by Eq.~\ref{actionMR}.
In this case, $S_N$ is a function of four unknown couplings, $s_{\emptyset,01},s_{22,\emptyset},s_{\emptyset,12}$ and $s_{3,01}$, and we minimise
	\eq{e_N = \left\lVert G_N(s_{\emptyset,01},s_{22,\emptyset},s_{\emptyset,12},s_{3,01})-O_N\right\rVert_F.}
We once again minimise this in the normalised basis of states at $\Delta L=4$.

In this case we have truncated the expansion at fourth order (once again, this is due to an explosion of extra coefficients at subsequent orders with three new parameters required at fifth order).
Therefore, we expect the errors on the coefficient $s_{\emptyset,01}$ to be $N^{-2}$ whereas for the two coefficients, $s_{22,\emptyset}$ and $s_{\emptyset,12}$, the errors are of order $N^{-1}$.
Finally, the error on $s_{3,01}$ is expected to be quite large; of order $\sqrt N^{-1}$.
As such, these coefficients are expected to have much larger errors than those in the Laughlin case.

We should also note that we have four coefficients as opposed to only two in the Laughlin case.
To some extent, this does make this description for the Moore-Read case less powerful as more numerical data is required.
However, the dimension of the space of edge states is also much larger in the Moore-Read case as it replicates exactly the edge states from the Laughlin state, made purely from bosonic modes, in addition to fermionic and mixed states.
For example, the $\Delta L=4$ sub-space has dimension 5 in the Laughlin case and 10 in the Moore-Read case.
There are therefore almost quadruple the number of distinct inner products one can consider at this angular momentum, which to some extent justifies the need for extra coefficients.

\begin{table*}[t!]
	\centering
	\begin{tabular}{| c || c | c | c | c | c |}
\hline $\nu = 1$ & $\Psi_{\bra{1,1,1\;;\;\emptyset}}$ & $\Psi_{\bra{2,1\;;\;\emptyset}}$ & $\Psi_{\bra{3\;;\;\emptyset}}$ & $\Psi_{\bra{1\;;\;\frac{1}{2},\frac{3}{2}}}$ & $\Psi_{\bra{\emptyset\;;\;\frac{1}{2},\frac{5}{2}}}$ \\ \hline\hline\multirow{2}{*}{$\Psi_{\bra{1,1,1\;;\;\emptyset}}$} & 6.0000\cellcolor{orange!20} & 0.3333 & 0.0185 & 0.0000 & 0.0000 \\ & \textbf{6.0000}\cellcolor{orange!20} & \textbf{0.3333} & \textbf{0.0185} & \textbf{0.0000} & \textbf{0.0000}\\ \hline\multirow{2}{*}{$\Psi_{\bra{2,1\;;\;\emptyset}}$} & 0.3333 & 1.9871\cellcolor{orange!20} & 0.3291 & 0.0600 & 0.0067 \\ & \textbf{0.3333} & \textbf{1.9877}\cellcolor{orange!20} & \textbf{0.3333} & \textbf{0.0593} & \textbf{0.0062}\\ \hline\multirow{2}{*}{$\Psi_{\bra{3\;;\;\emptyset}}$} & 0.0185 & 0.3291 & 2.8633\cellcolor{orange!20} & 0.0100 & 0.1924 \\ & \textbf{0.0185} & \textbf{0.3333} & \textbf{2.8660}\cellcolor{orange!20} & \textbf{0.0093} & \textbf{0.1895}\\ \hline\multirow{2}{*}{$\Psi_{\bra{1\;;\;\frac{1}{2},\frac{3}{2}}}$} & 0.0000 & 0.0600 & 0.0100 & 1.2916\cellcolor{orange!20} & 0.1435 \\ & \textbf{0.0000} & \textbf{0.0593} & \textbf{0.0093} & \textbf{1.2911}\cellcolor{orange!20} & \textbf{0.1440}\\ \hline\multirow{2}{*}{$\Psi_{\bra{\emptyset\;;\;\frac{1}{2},\frac{5}{2}}}$} & 0.0000 & 0.0067 & 0.1924 & 0.1435 & 1.4410\cellcolor{orange!20} \\ & \textbf{0.0000} & \textbf{0.0062} & \textbf{0.1895} & \textbf{0.1440} & \textbf{1.4413}\cellcolor{orange!20}\\ \hline
	\end{tabular}
	\caption{As in table \ref{5_2 overlaps} we show in row $i$ and column $j$ the inner product $\bbrakket{\Psi_{\bra{i}}}{\Psi_{\bra{j}}}/R^{2\Delta L}$ where the angular momentum of these edge excitations is $\Delta L=3$ and the system size is $N=18$.
		The states $\ket{i}$ are defined by Eq.~\ref{MR_edgestates}.
		This case is the Moore-Read state at filling $\nu=1$.
		The upper value in each cell is calculated by taking the overlaps exactly in the physical space of monomials whilst the lower value in bold is calculated by using $G_N$ using the extrapolations of $s_{\emptyset,01},s_{22,\emptyset},s_{\emptyset,12}$ and $s_{3,01}$ which we calculated in the previous section.}
	\label{MR3_1 overlaps}
\end{table*}

\begin{table*}[t!]
	\centering
	\begin{tabular}{| c || c | c | c | c | c |}
\hline $\nu = 1/2$ & $\Psi_{\bra{1,1,1\;;\;\emptyset}}$ & $\Psi_{\bra{2,1\;;\;\emptyset}}$ & $\Psi_{\bra{3\;;\;\emptyset}}$ & $\Psi_{\bra{1\;;\;\frac{1}{2},\frac{3}{2}}}$ & $\Psi_{\bra{\emptyset\;;\;\frac{1}{2},\frac{5}{2}}}$ \\ \hline\hline\multirow{2}{*}{$\Psi_{\bra{1,1,1\;;\;\emptyset}}$} & 6.0000\cellcolor{orange!20} & 0.2357 & 0.0093 & 0.0000 & 0.0000 \\ & \textbf{6.0000}\cellcolor{orange!20} & \textbf{0.2357} & \textbf{0.0093} & \textbf{0.0000} & \textbf{0.0000}\\ \hline\multirow{2}{*}{$\Psi_{\bra{2,1\;;\;\emptyset}}$} & 0.2357 & 1.9857\cellcolor{orange!20} & 0.2333 & 0.0436 & 0.0034 \\ & \textbf{0.2357} & \textbf{1.9857}\cellcolor{orange!20} & \textbf{0.2357} & \textbf{0.0408} & \textbf{0.0031}\\ \hline\multirow{2}{*}{$\Psi_{\bra{3\;;\;\emptyset}}$} & 0.0093 & 0.2333 & 2.8821\cellcolor{orange!20} & 0.0051 & 0.1291 \\ & \textbf{0.0093} & \textbf{0.2357} & \textbf{2.8838}\cellcolor{orange!20} & \textbf{0.0046} & \textbf{0.1273}\\ \hline\multirow{2}{*}{$\Psi_{\bra{1\;;\;\frac{1}{2},\frac{3}{2}}}$} & 0.0000 & 0.0436 & 0.0051 & 1.1556\cellcolor{orange!20} & 0.0908 \\ & \textbf{0.0000} & \textbf{0.0408} & \textbf{0.0046} & \textbf{1.1589}\cellcolor{orange!20} & \textbf{0.0917}\\ \hline\multirow{2}{*}{$\Psi_{\bra{\emptyset\;;\;\frac{1}{2},\frac{5}{2}}}$} & 0.0000 & 0.0034 & 0.1291 & 0.0908 & 1.2270\cellcolor{orange!20} \\ & \textbf{0.0000} & \textbf{0.0031} & \textbf{0.1273} & \textbf{0.0917} & \textbf{1.2183}\cellcolor{orange!20}\\ \hline
	\end{tabular}
	\caption{Similar to table \ref{MR3_1 overlaps} we show a comparison of the numerical calculations for the Moore-Read state (upper entry) with our approximation using $G_N$ (lower entry in bold) for $N=18$ and at filling fraction $\nu=1/2$.}
	\label{MR3_2 overlaps}
\end{table*}

Thus, we perform fits for these coefficients at fillings $\nu=1$ and $\nu=1/2$ and show some extrapolations as a function of $N$ in Fig.~\ref{MR fits} (once again, we show a selection of the data, $O_N$, used to fit these coefficients in the following subsection, in tables \ref{MR3_1 overlaps} and \ref{MR3_2 overlaps}, for the largest systems of $N=18$).
In the $\nu=1$ case each of the coefficients $s_{\emptyset,01},s_{22,\emptyset}$ and $s_{\emptyset,12}$ appear to obey the scaling hypothesis with small sub-leading corrections of order $\sqrt{N}^{-1}$.
Unfortunately, a similar extrapolation cannot be done for $s_{3,01}$ as the $\sqrt N^{-1}$ errors in the fits are too large.
However, we do see that the values are small (less than $0.01$ in each case, which is a similar size to the other coefficients) and so the effect of this term is not expected to be significant.
We also present the extrapolation of the least irrelevant coefficient $s_{\emptyset,01}$ at $\nu=1/2$.
Again, this appears to obey the scaling hypothesis, tending towards a constant value with a small sub-leading correction of order $\sqrt N^{-1}$.
However, we note that the sub-sub-leading corrections for this fermionic case appear to remain appreciable for the values of $N\le 18$ we present here.

\subsubsection{Accuracy of the $S_N$ expansion}

Using the fits for the coefficients as described above we can compare with the overlaps we find numerically.
In tables \ref{MR3_1 overlaps} and \ref{MR3_2 overlaps} we compare the exact overlaps at system sizes of $N=18$ with those calculated with our inner product operator $G_N$, as a function of the coefficient fits we found in the previous section.
We perform this comparison at $\Delta L=3$ and provide further data at $\Delta L=4$ in Appendix \ref{more data}.

We find that the agreement is once again very good, though unsurprisingly not quite as accurate as the higher-order calculation we performed for the Laughlin state.
The lack of a symmetry to constrain the coefficients of fermionic operators in $G_N$ also hampers the agreement for fermionic or mixed states, which are noticeably less accurate than their purely bosonic counterparts.

\section{Conclusion}

We have provided an in-depth analysis of the inner products of quantum Hall edge states in both the Laughlin and Moore-Read states.
We find that the form for the inner product conjectured in Ref.~\onlinecite{dubail2012edge} agrees very well with the numerical data, especially for larger system sizes.
We are also able to fit values of the coefficients involved in this expansion which cannot be constrained by symmetry by using exact methods, albeit at small system sizes.
The values determined for certain expansion coefficients match simple analytic forms which may suggest that additional physical constraints are present.
It would also be interesting to analyse the claims of Ref.~\onlinecite{dubail2012edge} in the context of further quantum Hall states, most notably the Read-Rezayi state.
Furthermore, we can examine these claims rigorously in the integer quantum Hall effect\cite{usFUTUREinteger} (which can be thought of as Laughlin at $\nu=1$) and apply a similar analysis to that considered here to inspect the dynamics of these edge modes\cite{usFUTUREhamiltonian}.

\section*{Acknowledgements}

We are grateful to J. Dubail for enlightening discussions.
This work was supported by EPSRC grants EP/I031014/1 and EP/N01930X/1.
Statement of compliance with EPSRC policy framework on research data: This publication is theoretical work that does not require supporting research data.

\bibliography{effective_FQH}

\begin{thebibliography}{33}%
\makeatletter
\providecommand \@ifxundefined [1]{%
 \@ifx{#1\undefined}
}%
\providecommand \@ifnum [1]{%
 \ifnum #1\expandafter \@firstoftwo
 \else \expandafter \@secondoftwo
 \fi
}%
\providecommand \@ifx [1]{%
 \ifx #1\expandafter \@firstoftwo
 \else \expandafter \@secondoftwo
 \fi
}%
\providecommand \natexlab [1]{#1}%
\providecommand \enquote  [1]{``#1''}%
\providecommand \bibnamefont  [1]{#1}%
\providecommand \bibfnamefont [1]{#1}%
\providecommand \citenamefont [1]{#1}%
\providecommand \href@noop [0]{\@secondoftwo}%
\providecommand \href [0]{\begingroup \@sanitize@url \@href}%
\providecommand \@href[1]{\@@startlink{#1}\@@href}%
\providecommand \@@href[1]{\endgroup#1\@@endlink}%
\providecommand \@sanitize@url [0]{\catcode `\\12\catcode `\$12\catcode
  `\&12\catcode `\#12\catcode `\^12\catcode `\_12\catcode `\%12\relax}%
\providecommand \@@startlink[1]{}%
\providecommand \@@endlink[0]{}%
\providecommand \url  [0]{\begingroup\@sanitize@url \@url }%
\providecommand \@url [1]{\endgroup\@href {#1}{\urlprefix }}%
\providecommand \urlprefix  [0]{URL }%
\providecommand \Eprint [0]{\href }%
\providecommand \doibase [0]{http://dx.doi.org/}%
\providecommand \selectlanguage [0]{\@gobble}%
\providecommand \bibinfo  [0]{\@secondoftwo}%
\providecommand \bibfield  [0]{\@secondoftwo}%
\providecommand \translation [1]{[#1]}%
\providecommand \BibitemOpen [0]{}%
\providecommand \bibitemStop [0]{}%
\providecommand \bibitemNoStop [0]{.\EOS\space}%
\providecommand \EOS [0]{\spacefactor3000\relax}%
\providecommand \BibitemShut  [1]{\csname bibitem#1\endcsname}%
\let\auto@bib@innerbib\@empty
\bibitem [{\citenamefont {Nayak}\ \emph {et~al.}(2008)\citenamefont {Nayak},
  \citenamefont {Simon}, \citenamefont {Stern}, \citenamefont {Freedman},\ and\
  \citenamefont {Sarma}}]{nayak2008non}%
  \BibitemOpen
  \bibfield  {author} {\bibinfo {author} {\bibfnamefont {C.}~\bibnamefont
  {Nayak}}, \bibinfo {author} {\bibfnamefont {S.~H.}\ \bibnamefont {Simon}},
  \bibinfo {author} {\bibfnamefont {A.}~\bibnamefont {Stern}}, \bibinfo
  {author} {\bibfnamefont {M.}~\bibnamefont {Freedman}}, \ and\ \bibinfo
  {author} {\bibfnamefont {S.~D.}\ \bibnamefont {Sarma}},\ }\href@noop {}
  {\bibfield  {journal} {\bibinfo  {journal} {Reviews of Modern Physics}\
  }\textbf {\bibinfo {volume} {80}},\ \bibinfo {pages} {1083} (\bibinfo {year}
  {2008})}\BibitemShut {NoStop}%
\bibitem [{\citenamefont {Girvin}(1999)}]{girvin1999quantum}%
  \BibitemOpen
  \bibfield  {author} {\bibinfo {author} {\bibfnamefont {S.~M.}\ \bibnamefont
  {Girvin}},\ }in\ \href@noop {} {\emph {\bibinfo {booktitle} {Aspects
  topologiques de la physique en basse dimension. Topological aspects of low
  dimensional systems}}}\ (\bibinfo  {publisher} {Springer},\ \bibinfo {year}
  {1999})\ pp.\ \bibinfo {pages} {53--175}\BibitemShut {NoStop}%
\bibitem [{\citenamefont {Moore}\ and\ \citenamefont
  {Read}(1991)}]{moore1991nonabelions}%
  \BibitemOpen
  \bibfield  {author} {\bibinfo {author} {\bibfnamefont {G.}~\bibnamefont
  {Moore}}\ and\ \bibinfo {author} {\bibfnamefont {N.}~\bibnamefont {Read}},\
  }\href@noop {} {\bibfield  {journal} {\bibinfo  {journal} {Nuclear Physics
  B}\ }\textbf {\bibinfo {volume} {360}},\ \bibinfo {pages} {362} (\bibinfo
  {year} {1991})}\BibitemShut {NoStop}%
\bibitem [{\citenamefont {Read}\ and\ \citenamefont
  {Rezayi}(1999)}]{read1999beyond}%
  \BibitemOpen
  \bibfield  {author} {\bibinfo {author} {\bibfnamefont {N.}~\bibnamefont
  {Read}}\ and\ \bibinfo {author} {\bibfnamefont {E.}~\bibnamefont {Rezayi}},\
  }\href@noop {} {\bibfield  {journal} {\bibinfo  {journal} {Physical Review
  B}\ }\textbf {\bibinfo {volume} {59}},\ \bibinfo {pages} {8084} (\bibinfo
  {year} {1999})}\BibitemShut {NoStop}%
\bibitem [{\citenamefont {Li}\ and\ \citenamefont
  {Haldane}(2008)}]{li2008entanglement}%
  \BibitemOpen
  \bibfield  {author} {\bibinfo {author} {\bibfnamefont {H.}~\bibnamefont
  {Li}}\ and\ \bibinfo {author} {\bibfnamefont {F.~D.~M.}\ \bibnamefont
  {Haldane}},\ }\href@noop {} {\bibfield  {journal} {\bibinfo  {journal}
  {Physical review letters}\ }\textbf {\bibinfo {volume} {101}},\ \bibinfo
  {pages} {010504} (\bibinfo {year} {2008})}\BibitemShut {NoStop}%
\bibitem [{\citenamefont {Dubail}\ \emph
  {et~al.}(2012{\natexlab{a}})\citenamefont {Dubail}, \citenamefont {Read},\
  and\ \citenamefont {Rezayi}}]{dubail2012edge}%
  \BibitemOpen
  \bibfield  {author} {\bibinfo {author} {\bibfnamefont {J.}~\bibnamefont
  {Dubail}}, \bibinfo {author} {\bibfnamefont {N.}~\bibnamefont {Read}}, \ and\
  \bibinfo {author} {\bibfnamefont {E.}~\bibnamefont {Rezayi}},\ }\href@noop {}
  {\bibfield  {journal} {\bibinfo  {journal} {Physical Review B}\ }\textbf
  {\bibinfo {volume} {86}},\ \bibinfo {pages} {245310} (\bibinfo {year}
  {2012}{\natexlab{a}})}\BibitemShut {NoStop}%
\bibitem [{\citenamefont {Hansson}\ \emph {et~al.}(2017)\citenamefont
  {Hansson}, \citenamefont {Hermanns}, \citenamefont {Simon},\ and\
  \citenamefont {Viefers}}]{hansson2017quantum}%
  \BibitemOpen
  \bibfield  {author} {\bibinfo {author} {\bibfnamefont {T.~H.}\ \bibnamefont
  {Hansson}}, \bibinfo {author} {\bibfnamefont {M.}~\bibnamefont {Hermanns}},
  \bibinfo {author} {\bibfnamefont {S.~H.}\ \bibnamefont {Simon}}, \ and\
  \bibinfo {author} {\bibfnamefont {S.~F.}\ \bibnamefont {Viefers}},\ }\href
  {\doibase 10.1103/RevModPhys.89.025005} {\bibfield  {journal} {\bibinfo
  {journal} {Rev. Mod. Phys.}\ }\textbf {\bibinfo {volume} {89}},\ \bibinfo
  {pages} {025005} (\bibinfo {year} {2017})}\BibitemShut {NoStop}%
\bibitem [{\citenamefont {Laughlin}(1983)}]{laughlin1983anomalous}%
  \BibitemOpen
  \bibfield  {author} {\bibinfo {author} {\bibfnamefont {R.~B.}\ \bibnamefont
  {Laughlin}},\ }\href@noop {} {\bibfield  {journal} {\bibinfo  {journal}
  {Physical Review Letters}\ }\textbf {\bibinfo {volume} {50}},\ \bibinfo
  {pages} {1395} (\bibinfo {year} {1983})}\BibitemShut {NoStop}%
\bibitem [{\citenamefont {Storni}\ \emph {et~al.}(2010)\citenamefont {Storni},
  \citenamefont {Morf},\ and\ \citenamefont {Sarma}}]{storni2010fractional}%
  \BibitemOpen
  \bibfield  {author} {\bibinfo {author} {\bibfnamefont {M.}~\bibnamefont
  {Storni}}, \bibinfo {author} {\bibfnamefont {R.}~\bibnamefont {Morf}}, \ and\
  \bibinfo {author} {\bibfnamefont {S.~D.}\ \bibnamefont {Sarma}},\ }\href@noop
  {} {\bibfield  {journal} {\bibinfo  {journal} {Physical review letters}\
  }\textbf {\bibinfo {volume} {104}},\ \bibinfo {pages} {076803} (\bibinfo
  {year} {2010})}\BibitemShut {NoStop}%
\bibitem [{\citenamefont {Rezayi}\ and\ \citenamefont
  {Read}(2009)}]{rezayi2009non}%
  \BibitemOpen
  \bibfield  {author} {\bibinfo {author} {\bibfnamefont {E.}~\bibnamefont
  {Rezayi}}\ and\ \bibinfo {author} {\bibfnamefont {N.}~\bibnamefont {Read}},\
  }\href@noop {} {\bibfield  {journal} {\bibinfo  {journal} {Physical Review
  B}\ }\textbf {\bibinfo {volume} {79}},\ \bibinfo {pages} {075306} (\bibinfo
  {year} {2009})}\BibitemShut {NoStop}%
\bibitem [{\citenamefont {Wen}(1990)}]{wen1990chiral}%
  \BibitemOpen
  \bibfield  {author} {\bibinfo {author} {\bibfnamefont {X.-G.}\ \bibnamefont
  {Wen}},\ }\href@noop {} {\bibfield  {journal} {\bibinfo  {journal} {Physical
  Review B}\ }\textbf {\bibinfo {volume} {41}},\ \bibinfo {pages} {12838}
  (\bibinfo {year} {1990})}\BibitemShut {NoStop}%
\bibitem [{\citenamefont {Chang}(2003)}]{chang2003chiral}%
  \BibitemOpen
  \bibfield  {author} {\bibinfo {author} {\bibfnamefont {A.}~\bibnamefont
  {Chang}},\ }\href@noop {} {\bibfield  {journal} {\bibinfo  {journal} {Reviews
  of Modern Physics}\ }\textbf {\bibinfo {volume} {75}},\ \bibinfo {pages}
  {1449} (\bibinfo {year} {2003})}\BibitemShut {NoStop}%
\bibitem [{\citenamefont {Wen}(1992)}]{wen1992theory}%
  \BibitemOpen
  \bibfield  {author} {\bibinfo {author} {\bibfnamefont {X.-G.}\ \bibnamefont
  {Wen}},\ }\href@noop {} {\bibfield  {journal} {\bibinfo  {journal}
  {International journal of modern physics B}\ }\textbf {\bibinfo {volume}
  {6}},\ \bibinfo {pages} {1711} (\bibinfo {year} {1992})}\BibitemShut
  {NoStop}%
\bibitem [{\citenamefont {Wiegmann}\ and\ \citenamefont
  {Zabrodin}(2005)}]{wiegmann2005large}%
  \BibitemOpen
  \bibfield  {author} {\bibinfo {author} {\bibfnamefont {P.~B.}\ \bibnamefont
  {Wiegmann}}\ and\ \bibinfo {author} {\bibfnamefont {A.}~\bibnamefont
  {Zabrodin}},\ }\href@noop {} {\bibfield  {journal} {\bibinfo  {journal} {J.
  Phys. A}\ }\textbf {\bibinfo {volume} {39}},\ \bibinfo {pages} {8933}
  (\bibinfo {year} {2005})}\BibitemShut {NoStop}%
\bibitem [{\citenamefont {Dubail}\ \emph
  {et~al.}(2012{\natexlab{b}})\citenamefont {Dubail}, \citenamefont {Read},\
  and\ \citenamefont {Rezayi}}]{dubail2012real}%
  \BibitemOpen
  \bibfield  {author} {\bibinfo {author} {\bibfnamefont {J.}~\bibnamefont
  {Dubail}}, \bibinfo {author} {\bibfnamefont {N.}~\bibnamefont {Read}}, \ and\
  \bibinfo {author} {\bibfnamefont {E.}~\bibnamefont {Rezayi}},\ }\href@noop {}
  {\bibfield  {journal} {\bibinfo  {journal} {Physical Review B}\ }\textbf
  {\bibinfo {volume} {85}},\ \bibinfo {pages} {115321} (\bibinfo {year}
  {2012}{\natexlab{b}})}\BibitemShut {NoStop}%
\bibitem [{\citenamefont {Fern}\ \emph {et~al.}({\natexlab{a}})\citenamefont
  {Fern}, \citenamefont {Bondesan},\ and\ \citenamefont
  {Simon}}]{usFUTUREinteger}%
  \BibitemOpen
  \bibfield  {author} {\bibinfo {author} {\bibfnamefont {R.}~\bibnamefont
  {Fern}}, \bibinfo {author} {\bibfnamefont {R.}~\bibnamefont {Bondesan}}, \
  and\ \bibinfo {author} {\bibfnamefont {S.}~\bibnamefont {Simon}},\
  }\href@noop {} {\emph {\bibinfo {title} {Future Publication}}}\BibitemShut
  {NoStop}%
\bibitem [{\citenamefont {Fern}\ \emph {et~al.}({\natexlab{b}})\citenamefont
  {Fern}, \citenamefont {Bondesan},\ and\ \citenamefont
  {Simon}}]{usFUTUREhamiltonian}%
  \BibitemOpen
  \bibfield  {author} {\bibinfo {author} {\bibfnamefont {R.}~\bibnamefont
  {Fern}}, \bibinfo {author} {\bibfnamefont {R.}~\bibnamefont {Bondesan}}, \
  and\ \bibinfo {author} {\bibfnamefont {S.}~\bibnamefont {Simon}},\
  }\href@noop {} {\emph {\bibinfo {title} {Future Publication}}}\BibitemShut
  {NoStop}%
\bibitem [{\citenamefont {Francesco}\ \emph {et~al.}(2012)\citenamefont
  {Francesco}, \citenamefont {Mathieu},\ and\ \citenamefont
  {S{\'e}n{\'e}chal}}]{francesco2012conformal}%
  \BibitemOpen
  \bibfield  {author} {\bibinfo {author} {\bibfnamefont {P.}~\bibnamefont
  {Francesco}}, \bibinfo {author} {\bibfnamefont {P.}~\bibnamefont {Mathieu}},
  \ and\ \bibinfo {author} {\bibfnamefont {D.}~\bibnamefont
  {S{\'e}n{\'e}chal}},\ }\href@noop {} {\emph {\bibinfo {title} {Conformal
  field theory}}}\ (\bibinfo  {publisher} {Springer Science \& Business
  Media},\ \bibinfo {year} {2012})\BibitemShut {NoStop}%
\bibitem [{\citenamefont {Ginsparg}(1988)}]{ginsparg1988applied}%
  \BibitemOpen
  \bibfield  {author} {\bibinfo {author} {\bibfnamefont {P.}~\bibnamefont
  {Ginsparg}},\ }\href@noop {} {\bibfield  {journal} {\bibinfo  {journal}
  {arXiv preprint hep-th/9108028}\ } (\bibinfo {year} {1988})}\BibitemShut
  {NoStop}%
\bibitem [{\citenamefont {Ribault}(2014)}]{ribault2014conformal}%
  \BibitemOpen
  \bibfield  {author} {\bibinfo {author} {\bibfnamefont {S.}~\bibnamefont
  {Ribault}},\ }\href@noop {} {\bibfield  {journal} {\bibinfo  {journal} {arXiv
  preprint arXiv:1406.4290}\ } (\bibinfo {year} {2014})}\BibitemShut {NoStop}%
\bibitem [{\citenamefont {Zibrov}\ \emph {et~al.}(2016)\citenamefont {Zibrov},
  \citenamefont {Kometter}, \citenamefont {Zhou}, \citenamefont {Spanton},
  \citenamefont {Taniguchi}, \citenamefont {Watanabe}, \citenamefont
  {Zaletel},\ and\ \citenamefont {Young}}]{zibrov2016robust}%
  \BibitemOpen
  \bibfield  {author} {\bibinfo {author} {\bibfnamefont {A.~A.}\ \bibnamefont
  {Zibrov}}, \bibinfo {author} {\bibfnamefont {C.}~\bibnamefont {Kometter}},
  \bibinfo {author} {\bibfnamefont {H.}~\bibnamefont {Zhou}}, \bibinfo {author}
  {\bibfnamefont {E.}~\bibnamefont {Spanton}}, \bibinfo {author} {\bibfnamefont
  {T.}~\bibnamefont {Taniguchi}}, \bibinfo {author} {\bibfnamefont
  {K.}~\bibnamefont {Watanabe}}, \bibinfo {author} {\bibfnamefont
  {M.}~\bibnamefont {Zaletel}}, \ and\ \bibinfo {author} {\bibfnamefont
  {A.}~\bibnamefont {Young}},\ }\href@noop {} {\bibfield  {journal} {\bibinfo
  {journal} {arXiv preprint arXiv:1611.07113}\ } (\bibinfo {year}
  {2016})}\BibitemShut {NoStop}%
\bibitem [{\citenamefont {Caillol}\ \emph {et~al.}(1982)\citenamefont
  {Caillol}, \citenamefont {Levesque}, \citenamefont {Weis},\ and\
  \citenamefont {Hansen}}]{caillol1982monte}%
  \BibitemOpen
  \bibfield  {author} {\bibinfo {author} {\bibfnamefont {J.}~\bibnamefont
  {Caillol}}, \bibinfo {author} {\bibfnamefont {D.}~\bibnamefont {Levesque}},
  \bibinfo {author} {\bibfnamefont {J.}~\bibnamefont {Weis}}, \ and\ \bibinfo
  {author} {\bibfnamefont {J.}~\bibnamefont {Hansen}},\ }\href@noop {}
  {\bibfield  {journal} {\bibinfo  {journal} {Journal of Statistical Physics}\
  }\textbf {\bibinfo {volume} {28}},\ \bibinfo {pages} {325} (\bibinfo {year}
  {1982})}\BibitemShut {NoStop}%
\bibitem [{\citenamefont {Bonderson}\ \emph {et~al.}(2011)\citenamefont
  {Bonderson}, \citenamefont {Gurarie},\ and\ \citenamefont
  {Nayak}}]{bonderson2011plasma}%
  \BibitemOpen
  \bibfield  {author} {\bibinfo {author} {\bibfnamefont {P.}~\bibnamefont
  {Bonderson}}, \bibinfo {author} {\bibfnamefont {V.}~\bibnamefont {Gurarie}},
  \ and\ \bibinfo {author} {\bibfnamefont {C.}~\bibnamefont {Nayak}},\
  }\href@noop {} {\bibfield  {journal} {\bibinfo  {journal} {Physical Review
  B}\ }\textbf {\bibinfo {volume} {83}},\ \bibinfo {pages} {075303} (\bibinfo
  {year} {2011})}\BibitemShut {NoStop}%
\bibitem [{\citenamefont {Read}(2009)}]{read2009non}%
  \BibitemOpen
  \bibfield  {author} {\bibinfo {author} {\bibfnamefont {N.}~\bibnamefont
  {Read}},\ }\href@noop {} {\bibfield  {journal} {\bibinfo  {journal} {Physical
  Review B}\ }\textbf {\bibinfo {volume} {79}},\ \bibinfo {pages} {045308}
  (\bibinfo {year} {2009})}\BibitemShut {NoStop}%
\bibitem [{\citenamefont {Girvin}\ and\ \citenamefont
  {Jach}(1984)}]{girvin1984formalism}%
  \BibitemOpen
  \bibfield  {author} {\bibinfo {author} {\bibfnamefont {S.}~\bibnamefont
  {Girvin}}\ and\ \bibinfo {author} {\bibfnamefont {T.}~\bibnamefont {Jach}},\
  }\href@noop {} {\bibfield  {journal} {\bibinfo  {journal} {Physical Review
  B}\ }\textbf {\bibinfo {volume} {29}},\ \bibinfo {pages} {5617} (\bibinfo
  {year} {1984})}\BibitemShut {NoStop}%
\bibitem [{\citenamefont {Bernevig}\ and\ \citenamefont
  {Haldane}(2008{\natexlab{a}})}]{bernevig2008generalized}%
  \BibitemOpen
  \bibfield  {author} {\bibinfo {author} {\bibfnamefont {B.~A.}\ \bibnamefont
  {Bernevig}}\ and\ \bibinfo {author} {\bibfnamefont {F.}~\bibnamefont
  {Haldane}},\ }\href@noop {} {\bibfield  {journal} {\bibinfo  {journal}
  {Physical Review B}\ }\textbf {\bibinfo {volume} {77}},\ \bibinfo {pages}
  {184502} (\bibinfo {year} {2008}{\natexlab{a}})}\BibitemShut {NoStop}%
\bibitem [{\citenamefont {Bernevig}\ and\ \citenamefont
  {Haldane}(2008{\natexlab{b}})}]{bernevig2008model}%
  \BibitemOpen
  \bibfield  {author} {\bibinfo {author} {\bibfnamefont {B.~A.}\ \bibnamefont
  {Bernevig}}\ and\ \bibinfo {author} {\bibfnamefont {F.}~\bibnamefont
  {Haldane}},\ }\href@noop {} {\bibfield  {journal} {\bibinfo  {journal}
  {Physical review letters}\ }\textbf {\bibinfo {volume} {100}},\ \bibinfo
  {pages} {246802} (\bibinfo {year} {2008}{\natexlab{b}})}\BibitemShut
  {NoStop}%
\bibitem [{\citenamefont {Bernevig}\ and\ \citenamefont
  {Haldane}(2008{\natexlab{c}})}]{bernevig2008properties}%
  \BibitemOpen
  \bibfield  {author} {\bibinfo {author} {\bibfnamefont {B.~A.}\ \bibnamefont
  {Bernevig}}\ and\ \bibinfo {author} {\bibfnamefont {F.}~\bibnamefont
  {Haldane}},\ }\href@noop {} {\bibfield  {journal} {\bibinfo  {journal}
  {Physical review letters}\ }\textbf {\bibinfo {volume} {101}},\ \bibinfo
  {pages} {246806} (\bibinfo {year} {2008}{\natexlab{c}})}\BibitemShut
  {NoStop}%
\bibitem [{\citenamefont {Baratta}\ and\ \citenamefont
  {Forrester}(2011)}]{baratta2011jack}%
  \BibitemOpen
  \bibfield  {author} {\bibinfo {author} {\bibfnamefont {W.}~\bibnamefont
  {Baratta}}\ and\ \bibinfo {author} {\bibfnamefont {P.~J.}\ \bibnamefont
  {Forrester}},\ }\href@noop {} {\bibfield  {journal} {\bibinfo  {journal}
  {Nuclear Physics B}\ }\textbf {\bibinfo {volume} {843}},\ \bibinfo {pages}
  {362} (\bibinfo {year} {2011})}\BibitemShut {NoStop}%
\bibitem [{\citenamefont {Zaletel}\ and\ \citenamefont
  {Mong}(2012)}]{zaletel2012exact}%
  \BibitemOpen
  \bibfield  {author} {\bibinfo {author} {\bibfnamefont {M.~P.}\ \bibnamefont
  {Zaletel}}\ and\ \bibinfo {author} {\bibfnamefont {R.~S.}\ \bibnamefont
  {Mong}},\ }\href@noop {} {\bibfield  {journal} {\bibinfo  {journal} {Physical
  Review B}\ }\textbf {\bibinfo {volume} {86}},\ \bibinfo {pages} {245305}
  (\bibinfo {year} {2012})}\BibitemShut {NoStop}%
\bibitem [{\citenamefont {Estienne}\ \emph {et~al.}(2013)\citenamefont
  {Estienne}, \citenamefont {Papi{\'c}}, \citenamefont {Regnault},\ and\
  \citenamefont {Bernevig}}]{estienne2013matrix}%
  \BibitemOpen
  \bibfield  {author} {\bibinfo {author} {\bibfnamefont {B.}~\bibnamefont
  {Estienne}}, \bibinfo {author} {\bibfnamefont {Z.}~\bibnamefont {Papi{\'c}}},
  \bibinfo {author} {\bibfnamefont {N.}~\bibnamefont {Regnault}}, \ and\
  \bibinfo {author} {\bibfnamefont {B.~A.}\ \bibnamefont {Bernevig}},\
  }\href@noop {} {\bibfield  {journal} {\bibinfo  {journal} {Physical Review
  B}\ }\textbf {\bibinfo {volume} {87}},\ \bibinfo {pages} {161112} (\bibinfo
  {year} {2013})}\BibitemShut {NoStop}%
\bibitem [{\citenamefont {Abanov}\ and\ \citenamefont
  {Wiegmann}(2005)}]{abanov2005quantum}%
  \BibitemOpen
  \bibfield  {author} {\bibinfo {author} {\bibfnamefont {A.~G.}\ \bibnamefont
  {Abanov}}\ and\ \bibinfo {author} {\bibfnamefont {P.~B.}\ \bibnamefont
  {Wiegmann}},\ }\href@noop {} {\bibfield  {journal} {\bibinfo  {journal}
  {Physical review letters}\ }\textbf {\bibinfo {volume} {95}},\ \bibinfo
  {pages} {076402} (\bibinfo {year} {2005})}\BibitemShut {NoStop}%
\bibitem [{\citenamefont {Wiegmann}(2012)}]{wiegmann2012nonlinear}%
  \BibitemOpen
  \bibfield  {author} {\bibinfo {author} {\bibfnamefont {P.}~\bibnamefont
  {Wiegmann}},\ }\href@noop {} {\bibfield  {journal} {\bibinfo  {journal}
  {Physical review letters}\ }\textbf {\bibinfo {volume} {108}},\ \bibinfo
  {pages} {206810} (\bibinfo {year} {2012})}\BibitemShut {NoStop}%
\end{thebibliography}%

\FloatBarrier

\appendix

\section{A Linearly Independent Basis}
\label{linear_ind_app}

At any given scaling dimension we are interested in finding a linearly independent basis of terms which might appear in the action.
Let us consider this for the Laughlin case where,
	\eq{T_\bgam = \oint\cint{z}z^{d_a-1}\prod_{n\in\bgam}i\partial^n\varphi(z).}
Therefore, any contribution appearing at scaling dimension $d_a-1$ can be written as
	\eq{\mathcal{T}_{d_a-1} = \sum_{|\bgam|=d_a}s_\bgam T_\bgam}
where this sum is over all partitions $\bgam=\{\Gamma_1,\Gamma_2,\hdots\}$ which sum to $d_a$ (these partitions constitute all the ways of differentiating the field such that the total number of derivatives is $d_a$).
Certain $T_\bgam$ can then be used as new, linearly independent operators in addition to the basis of operators which arise from lower scaling dimensions.

Let us see how this works in practice.
As mentioned in the text, the $d_a=1$ sector is somewhat trivial.
It does give us $T_1$, but this operator is null ($T_1=a_0=0$).
Therefore, the first non-trivial case arises at $d_a=2$.
Here, there are two partitions, $T_2$ and $T_{11}$.
The former can be integrated by parts and gives us back our null operator.
The latter, however, is distinct, containing two bosonic modes.
Thus, at scaling dimensions $d_a= 2$ a full, linearly independent basis is given by
	\eq{T_\bgam \in \{T_{11}\}.}

We then move onto $d_a=3$, where the possibilities are $T_{3}$, $T_{21}$ and $T_{111}$.
Once again, the first case is a couple of integrations by parts away from our null operator, $T_1$.
The last case, $T_{111}$, is a distinct operator, containing three bosonic modes, and so indescribable by a linear combination of any other terms which are all two-bosons or fewer.
Finally, the middle case, $T_{21}$ can be integrated by parts to give
	\eq{T_{21} = - T_{11}.}
Therefore, this is not new.
Our linearly independent basis of operators for $d_a= 3$ is
	\eq{T_\bgam \in \{T_{11}, T_{111}\}.}
	
Going forward, the view is much the same.
We continue to add partitions, $\bgam$, of higher scaling dimension, $d_a$, and attempt to phrase them in terms of basis operators we already have.
To do so, we will try to reduce the size of the maximal derivatives in that term, i.e, given some term $T_{311}$ our first priority should be to integrate away the $3$.

So consider the next level, $d_a=4$, where we have
	\alg{T_4 & = - 3\cancel{T_3}, \\
		T_{31} & = - 3\cancel{T_{21}} - T_{22}, \\
		T_{22}, & \\
		T_{211} & = - \cancel{T_{111}}, \\
		T_{1111}. &}
We have crossed out any terms with scaling dimension $d_a<4$, as we've already considered whether such terms produce linearly independent basis functions.
We note that we are left with only two operators we cannot reduce in this way, $T_{22}$ and $T_{1111}$.
Once again, it should be clear that $T_{1111}$ is independent as it is a four-boson term, and our current basis, $\{T_{11}, T_{111}\}$, doesn't contain any other four-boson terms.

However, could $T_{22}$ simply be $T_{11}$ in disguise?
Whilst we cannot massage one into the other via some integration by parts procedure, they may well describe the same matrix within our basis of edge states and therefore it is necessary to check that this is not the case.
For this case it is simple as both are diagonal.
In fact, in the $\Delta L=2$ subspace (spanned by $\ket{1,1}$ and $\ket{2}$) they have the forms
	\eq{T_{11} = \begin{pmatrix}4&0\\0&4\end{pmatrix}\qquad\qquad
	T_{22} = \begin{pmatrix}0&0\\0&12\end{pmatrix}.}
These are clearly independent matrices.
Thus, for $d_a= 4$ the independent set of operators is
	\eq{T_\bgam \in \{T_{11}, T_{111}, T_{1111}, T_{22}\}.}

Let us conclude with one final example, at scaling dimension $d_a=5$.
The terms here are
	\alg{T_5 & = -4\cancel{T_4} \\
		T_{41} & = -4\cancel{T_{31}} - T_{32} \\
		T_{32} & = -2\cancel{T_{22}} \\
		T_{311} & = -4\cancel{T_{211}} - 2T_{221} \\
		T_{221} & \\
		T_{2111} & = -\cancel{T_{1111}} \\
		T_{11111}, &}
leaving at most two new basis terms, $T_{221}$ and $T_{11111}$.
Once again, analysing individual matrix elements convinces us that both of these are new and independent terms, so for $d_a= 5$,
	\eq{T_\bgam \in \{T_{11},T_{111},T_{1111},T_{22},T_{11111},T_{221}\}.}

In general, an efficient method for finding unique $\bgam$ is one of reducing the terms $T_\bgam$ to terms with at least the first two integers $\Gamma_1$ and $\Gamma_2$ being equal.
Such terms are difficult to reduce further integrating by parts so one must see if the resulting terms are linearly independent by other means, for example by considering individual matrix elements.

\section{Constraints on non-local terms}
\label{non local}

Throughout this work we have assumed that the action, $S_N$ contains no non-local terms.
This claim is based on the premise that the Laughlin state is in the screening phase, and so has short-range correlations within the bulk.
As such, this perturbation to the underlying CFT, our action in $G_N$, should depend only on local degrees of freedom.

It is interesting now to assess this claim in light of the data.
Consider some new action of the form
	\eq{\tilde S_N = S_N + S_N^\tr{(non-local)}}
where this extra contribution now contains anything which might be non-local.
The simplest such terms which can couple the field at two arbitrarily separated positions are those with two integrals.
The most well known of these which also respects our rotational invariance condition is the Benjamin-Ono term\cite{abanov2005quantum,wiegmann2012nonlinear}, of the form
	\alg{T_\tr{B-O} & = \underset{|z|>|w|}{\oint\cint{z}\oint\cint{w}}\frac{zw}{(z-w)^2}:i\partial\varphi(z)i\partial\varphi(w):
				\nonumber \\&
				= \sum_{n>0}na_{-n}a_n.}
The coefficient of this term is expected to scale as $N^{-1}$.

However, this Benjamin-Ono term by itself does not respect translational invariance and therefore, to consider whether it appears or not, we create a new translationally invariant Benjamin-Ono term by combining it with $T_{11}$, such that the result commutes with $a_{-1}$.
This has the form
	\eq{\tilde T_\tr{B-O} = \sum_{n>0}(n-1)a_{-n}a_n.}
It is then possible for this term to appear in the expansion of $S_N$ but with some unknown coefficient, $\frac{s_\tr{BO}}{N}$, which cannot be fixed by translational symmetry.
Therefore, if we include only this term in $S_N^\tr{(non-local)}$ for simplicity, we have one extra parameter we must fit.

In this vein we expand our new $\tilde G_N$, whose action is $\tilde S_N$, and attempt to fit this new coefficient.
We take an $N^{-2}$ expansion, neglecting any terms of order $N^{-5/2}$, and fit the coefficients, $s_{22}$ and $s_\tr{BO}$ by using the $\bbrakket{\Psi_{\bra{0}a_2}}{\Psi_{\bra{0}a_2}}$ and $\bbrakket{\Psi_{\bra{0}a_3}}{\Psi_{\bra{0}a_3}}$ overlaps for calibration.
I.e, we choose $s_{22}$ and $s_\tr{BO}$ such that the numerically calculated normalisations of these two states agree exactly with our expression, $\tilde G_N$.

\begin{figure}[t!]
	\centering
	\includegraphics[scale=0.43]{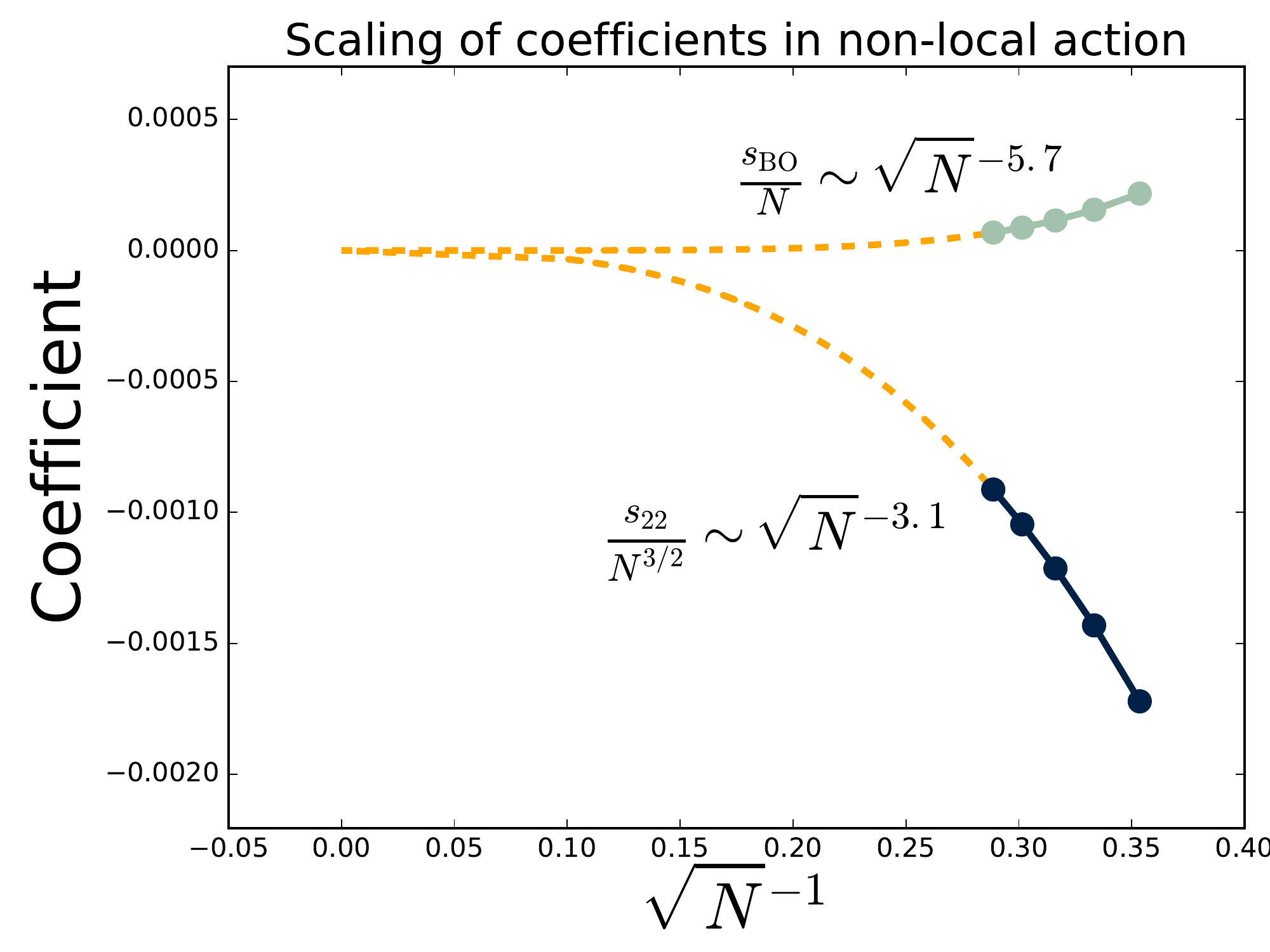}
	\caption{Na\"ive power law scaling of the coefficients of the Benjamin-Ono term (light blue) and the $T_{22}$ term (dark blue) for $\nu=1/2$ Laughlin when we expand the non-local metric, $\tilde G_N$, neglecting terms of order $N^{-5/2}$.
		We find that $s_{22}$ scales as expected $\sim N^{-3/2}$ whilst the Benjamin-Ono term falls off faster than $N^{-5/2}$, which is the order of terms we neglect.
		This analysis does not rule out the existence of all non-local terms but is further evidence as to their insignificance.}
	\label{B-O}
\end{figure}

The result of the fit is shown in Fig.~\ref{B-O}.
It shows that the coefficient $s_{22}$ scales approximately as expected, appearing very close to $3^\tr{rd}$ order in the expansion, as corresponding to the scaling dimension of the $T_{22}$ term.
On the other hand, the Benjamin-Ono term falls of roughly as $N^{-5/2}$ or faster, which is the order of terms we are neglecting in this calculation.
Therefore, this fit is as likely to come from the corrections we are neglecting as it is to be due to a non-zero Benjamin-Ono coefficient.
This gives additional evidence supporting the notion that all terms should be local.

\section{Further Data}
\label{more data}

In tables \ref{MR4_1 overlaps} and \ref{MR4_2 overlaps} we show further data at higher angular momenta for the Moore-Read states at filling $\nu=1$ and $\nu=1/2$.
In both cases the agreement between exact numerics and the effective description using $G_N$ is shown for edge states with angular momentum $\Delta L=4$.
In both cases we find that this agreement is very good.

\def\arraystretch{1.5}

\begin{sidewaystable*}[h!]
	\vspace{9cm}
	\centering
	\begin{tabular}{| c || c | c | c | c | c | c | c | c | c | c |}
\hline $\nu = 1$ & $\Psi_{\bra{1,1,1,1\;;\;\emptyset}}$ & $\Psi_{\bra{2,1,1\;;\;\emptyset}}$ & $\Psi_{\bra{2,2\;;\;\emptyset}}$ & $\Psi_{\bra{3,1\;;\;\emptyset}}$ & $\Psi_{\bra{4\;;\;\emptyset}}$ & $\Psi_{\bra{1,1\;;\;\frac{1}{2},\frac{3}{2}}}$ & $\Psi_{\bra{2\;;\;\frac{1}{2},\frac{3}{2}}}$ & $\Psi_{\bra{1\;;\;\frac{1}{2},\frac{5}{2}}}$ & $\Psi_{\bra{\emptyset\;;\;\frac{1}{2},\frac{7}{2}}}$ & $\Psi_{\bra{\emptyset\;;\;\frac{3}{2},\frac{5}{2}}}$ \\ \hline\hline\multirow{2}{*}{$\Psi_{\bra{1,1,1,1\;;\;\emptyset}}$} & 24.0000\cellcolor{orange!20} & 1.3333 & 0.0741 & 0.0741 & 0.0041 & 0.0000 & 0.0000 & 0.0000 & 0.0000 & 0.0000 \\ & \textbf{24.0000}\cellcolor{orange!20} & \textbf{1.3333} & \textbf{0.0741} & \textbf{0.0741} & \textbf{0.0000} & \textbf{0.0000} & \textbf{0.0000} & \textbf{0.0000} & \textbf{0.0000} & \textbf{0.0000}\\ \hline\multirow{2}{*}{$\Psi_{\bra{2,1,1\;;\;\emptyset}}$} & 1.3333 & 4.0113\cellcolor{orange!20} & 0.4416 & 0.6603 & 0.0731 & 0.1199 & 0.0067 & 0.0133 & 0.0011 & 0.0004 \\ & \textbf{1.3333} & \textbf{4.0124}\cellcolor{orange!20} & \textbf{0.4444} & \textbf{0.6667} & \textbf{0.0741} & \textbf{0.1185} & \textbf{0.0062} & \textbf{0.0123} & \textbf{0.0000} & \textbf{0.0000}\\ \hline\multirow{2}{*}{$\Psi_{\bra{2,2\;;\;\emptyset}}$} & 0.0741 & 0.4416 & 7.8482\cellcolor{orange!20} & 0.0731 & 0.8546 & 0.0133 & 0.2382 & 0.0015 & 0.0387 & -0.0039 \\ & \textbf{0.0741} & \textbf{0.4444} & \textbf{7.8522}\cellcolor{orange!20} & \textbf{0.0741} & \textbf{0.8889} & \textbf{0.0123} & \textbf{0.2371} & \textbf{0.0000} & \textbf{0.0340} & \textbf{-0.0031}\\ \hline\multirow{2}{*}{$\Psi_{\bra{3,1\;;\;\emptyset}}$} & 0.0741 & 0.6603 & 0.0731 & 2.9181\cellcolor{orange!20} & 0.6363 & 0.0200 & 0.0011 & 0.1935 & 0.0321 & 0.0107 \\ & \textbf{0.0741} & \textbf{0.6667} & \textbf{0.0741} & \textbf{2.9216}\cellcolor{orange!20} & \textbf{0.6667} & \textbf{0.0185} & \textbf{0.0000} & \textbf{0.1895} & \textbf{0.0278} & \textbf{0.0093}\\ \hline\multirow{2}{*}{$\Psi_{\bra{4\;;\;\emptyset}}$} & 0.0041 & 0.0731 & 0.8546 & 0.6363 & 3.6003\cellcolor{orange!20} & 0.0022 & 0.0420 & 0.0428 & 0.4056 & 0.1369 \\ & \textbf{0.0000} & \textbf{0.0741} & \textbf{0.8889} & \textbf{0.6667} & \textbf{3.5997}\cellcolor{orange!20} & \textbf{0.0000} & \textbf{0.0370} & \textbf{0.0370} & \textbf{0.4037} & \textbf{0.1346}\\ \hline\multirow{2}{*}{$\Psi_{\bra{1,1\;;\;\frac{1}{2},\frac{3}{2}}}$} & 0.0000 & 0.1199 & 0.0133 & 0.0200 & 0.0022 & 2.5832\cellcolor{orange!20} & 0.1435 & 0.2870 & 0.0239 & 0.0080 \\ & \textbf{0.0000} & \textbf{0.1185} & \textbf{0.0123} & \textbf{0.0185} & \textbf{0.0000} & \textbf{2.5822}\cellcolor{orange!20} & \textbf{0.1440} & \textbf{0.2881} & \textbf{0.0185} & \textbf{0.0062}\\ \hline\multirow{2}{*}{$\Psi_{\bra{2\;;\;\frac{1}{2},\frac{3}{2}}}$} & 0.0000 & 0.0067 & 0.2382 & 0.0011 & 0.0420 & 0.1435 & 2.5856\cellcolor{orange!20} & 0.0159 & 0.3803 & -0.2304 \\ & \textbf{0.0000} & \textbf{0.0062} & \textbf{0.2371} & \textbf{0.0000} & \textbf{0.0370} & \textbf{0.1440} & \textbf{2.5777}\cellcolor{orange!20} & \textbf{0.0123} & \textbf{0.3834} & \textbf{-0.2300}\\ \hline\multirow{2}{*}{$\Psi_{\bra{1\;;\;\frac{1}{2},\frac{5}{2}}}$} & 0.0000 & 0.0133 & 0.0015 & 0.1935 & 0.0428 & 0.2870 & 0.0159 & 1.4570\cellcolor{orange!20} & 0.2402 & 0.0801 \\ & \textbf{0.0000} & \textbf{0.0123} & \textbf{0.0000} & \textbf{0.1895} & \textbf{0.0370} & \textbf{0.2881} & \textbf{0.0123} & \textbf{1.4537}\cellcolor{orange!20} & \textbf{0.2451} & \textbf{0.0817}\\ \hline\multirow{2}{*}{$\Psi_{\bra{\emptyset\;;\;\frac{1}{2},\frac{7}{2}}}$} & 0.0000 & 0.0011 & 0.0387 & 0.0321 & 0.4056 & 0.0239 & 0.3803 & 0.2402 & 1.5792\cellcolor{orange!20} & -0.0032 \\ & \textbf{0.0000} & \textbf{0.0000} & \textbf{0.0340} & \textbf{0.0278} & \textbf{0.4037} & \textbf{0.0185} & \textbf{0.3834} & \textbf{0.2451} & \textbf{1.5797}\cellcolor{orange!20} & \textbf{-0.0023}\\ \hline\multirow{2}{*}{$\Psi_{\bra{\emptyset\;;\;\frac{3}{2},\frac{5}{2}}}$} & 0.0000 & 0.0004 & -0.0039 & 0.0107 & 0.1369 & 0.0080 & -0.2304 & 0.0801 & -0.0032 & 1.6294\cellcolor{orange!20} \\ & \textbf{0.0000} & \textbf{0.0000} & \textbf{-0.0031} & \textbf{0.0093} & \textbf{0.1346} & \textbf{0.0062} & \textbf{-0.2300} & \textbf{0.0817} & \textbf{-0.0023} & \textbf{1.6318}\cellcolor{orange!20}\\ \hline
	\end{tabular}
	\caption{A table showing the agreement between the exactly calculated overlaps of quantum Hall edge wavefunctions in the Moore-Read state with angular momentum $\Delta L=4$ and the form of the same overlaps given by $G_N$.
		The upper entries are the exact data whereas the lower (bold) entries are the fit given by $G_N(s_{\emptyset,01},s_{22,\emptyset},s_{\emptyset,12},s_{3,01})$.}
	\label{MR4_1 overlaps}
\end{sidewaystable*}

\begin{sidewaystable*}[t!]
	\vspace{9cm}
	\centering
	\begin{tabular}{| c || c | c | c | c | c | c | c | c | c | c |}
\hline $\nu = 1/2$ & $\Psi_{\bra{1,1,1,1\;;\;\emptyset}}$ & $\Psi_{\bra{2,1,1\;;\;\emptyset}}$ & $\Psi_{\bra{2,2\;;\;\emptyset}}$ & $\Psi_{\bra{3,1\;;\;\emptyset}}$ & $\Psi_{\bra{4\;;\;\emptyset}}$ & $\Psi_{\bra{1,1\;;\;\frac{1}{2},\frac{3}{2}}}$ & $\Psi_{\bra{2\;;\;\frac{1}{2},\frac{3}{2}}}$ & $\Psi_{\bra{1\;;\;\frac{1}{2},\frac{5}{2}}}$ & $\Psi_{\bra{\emptyset\;;\;\frac{1}{2},\frac{7}{2}}}$ & $\Psi_{\bra{\emptyset\;;\;\frac{3}{2},\frac{5}{2}}}$ \\ \hline\hline\multirow{2}{*}{$\Psi_{\bra{1,1,1,1\;;\;\emptyset}}$} & 24.0000\cellcolor{orange!20} & 0.9428 & 0.0370 & 0.0370 & 0.0015 & 0.0000 & 0.0000 & 0.0000 & 0.0000 & 0.0000 \\ & \textbf{24.0000}\cellcolor{orange!20} & \textbf{0.9428} & \textbf{0.0370} & \textbf{0.0370} & \textbf{0.0000} & \textbf{0.0000} & \textbf{0.0000} & \textbf{0.0000} & \textbf{0.0000} & \textbf{0.0000}\\ \hline\multirow{2}{*}{$\Psi_{\bra{2,1,1\;;\;\emptyset}}$} & 0.9428 & 3.9899\cellcolor{orange!20} & 0.3120 & 0.4673 & 0.0367 & 0.0871 & 0.0034 & 0.0068 & 0.0004 & 0.0001 \\ & \textbf{0.9428} & \textbf{3.9898}\cellcolor{orange!20} & \textbf{0.3143} & \textbf{0.4714} & \textbf{0.0370} & \textbf{0.0817} & \textbf{0.0031} & \textbf{0.0062} & \textbf{0.0000} & \textbf{0.0000}\\ \hline\multirow{2}{*}{$\Psi_{\bra{2,2\;;\;\emptyset}}$} & 0.0370 & 0.3120 & 7.8612\cellcolor{orange!20} & 0.0367 & 0.6075 & 0.0068 & 0.1714 & 0.0005 & 0.0203 & -0.0035 \\ & \textbf{0.0370} & \textbf{0.3143} & \textbf{7.8606}\cellcolor{orange!20} & \textbf{0.0370} & \textbf{0.6285} & \textbf{0.0062} & \textbf{0.1633} & \textbf{0.0000} & \textbf{0.0170} & \textbf{-0.0015}\\ \hline\multirow{2}{*}{$\Psi_{\bra{3,1\;;\;\emptyset}}$} & 0.0370 & 0.4673 & 0.0367 & 2.9096\cellcolor{orange!20} & 0.4529 & 0.0103 & 0.0004 & 0.1295 & 0.0152 & 0.0051 \\ & \textbf{0.0370} & \textbf{0.4714} & \textbf{0.0370} & \textbf{2.9116}\cellcolor{orange!20} & \textbf{0.4714} & \textbf{0.0093} & \textbf{0.0000} & \textbf{0.1273} & \textbf{0.0139} & \textbf{0.0046}\\ \hline\multirow{2}{*}{$\Psi_{\bra{4\;;\;\emptyset}}$} & 0.0015 & 0.0367 & 0.6075 & 0.4529 & 3.6354\cellcolor{orange!20} & 0.0008 & 0.0231 & 0.0203 & 0.2743 & 0.0857 \\ & \textbf{0.0000} & \textbf{0.0370} & \textbf{0.6285} & \textbf{0.4714} & \textbf{3.6360}\cellcolor{orange!20} & \textbf{0.0000} & \textbf{0.0185} & \textbf{0.0185} & \textbf{0.2645} & \textbf{0.0882}\\ \hline\multirow{2}{*}{$\Psi_{\bra{1,1\;;\;\frac{1}{2},\frac{3}{2}}}$} & 0.0000 & 0.0871 & 0.0068 & 0.0103 & 0.0008 & 2.3112\cellcolor{orange!20} & 0.0908 & 0.1816 & 0.0107 & 0.0036 \\ & \textbf{0.0000} & \textbf{0.0817} & \textbf{0.0062} & \textbf{0.0093} & \textbf{0.0000} & \textbf{2.3177}\cellcolor{orange!20} & \textbf{0.0917} & \textbf{0.1834} & \textbf{0.0093} & \textbf{0.0031}\\ \hline\multirow{2}{*}{$\Psi_{\bra{2\;;\;\frac{1}{2},\frac{3}{2}}}$} & 0.0000 & 0.0034 & 0.1714 & 0.0004 & 0.0231 & 0.0908 & 2.3035\cellcolor{orange!20} & 0.0071 & 0.2382 & -0.1395 \\ & \textbf{0.0000} & \textbf{0.0031} & \textbf{0.1633} & \textbf{0.0000} & \textbf{0.0185} & \textbf{0.0917} & \textbf{2.3075}\cellcolor{orange!20} & \textbf{0.0062} & \textbf{0.2381} & \textbf{-0.1429}\\ \hline\multirow{2}{*}{$\Psi_{\bra{1\;;\;\frac{1}{2},\frac{5}{2}}}$} & 0.0000 & 0.0068 & 0.0005 & 0.1295 & 0.0203 & 0.1816 & 0.0071 & 1.2342\cellcolor{orange!20} & 0.1446 & 0.0482 \\ & \textbf{0.0000} & \textbf{0.0062} & \textbf{0.0000} & \textbf{0.1273} & \textbf{0.0185} & \textbf{0.1834} & \textbf{0.0062} & \textbf{1.2245}\cellcolor{orange!20} & \textbf{0.1484} & \textbf{0.0495}\\ \hline\multirow{2}{*}{$\Psi_{\bra{\emptyset\;;\;\frac{1}{2},\frac{7}{2}}}$} & 0.0000 & 0.0004 & 0.0203 & 0.0152 & 0.2743 & 0.0107 & 0.2382 & 0.1446 & 1.2474\cellcolor{orange!20} & -0.0015 \\ & \textbf{0.0000} & \textbf{0.0000} & \textbf{0.0170} & \textbf{0.0139} & \textbf{0.2645} & \textbf{0.0093} & \textbf{0.2381} & \textbf{0.1484} & \textbf{1.2417}\cellcolor{orange!20} & \textbf{-0.0012}\\ \hline\multirow{2}{*}{$\Psi_{\bra{\emptyset\;;\;\frac{3}{2},\frac{5}{2}}}$} & 0.0000 & 0.0001 & -0.0035 & 0.0051 & 0.0857 & 0.0036 & -0.1395 & 0.0482 & -0.0015 & 1.3021\cellcolor{orange!20} \\ & \textbf{0.0000} & \textbf{0.0000} & \textbf{-0.0015} & \textbf{0.0046} & \textbf{0.0882} & \textbf{0.0031} & \textbf{-0.1429} & \textbf{0.0495} & \textbf{-0.0012} & \textbf{1.3061}\cellcolor{orange!20}\\ \hline
	\end{tabular}
	\caption{As in Fig.~\ref{MR4_2 overlaps} we compare the exactly calculated Moore-Read state overlaps (upper) with the values calculated using $G_N$ (lower, bold).
		Here we consider the $\nu=1/2$ case and once again find very good agreement.}
	\label{MR4_2 overlaps}
\end{sidewaystable*}

\end{document}